\documentclass[12pt,reqno]{amsart} % add titlepage param for separate title page

\usepackage[margin=1in]{geometry}
\usepackage{blindtext}
\usepackage{graphicx}
\usepackage{subcaption}

\usepackage{mathtools,
  booktabs,
  braket,
  mathrsfs,
  latexsym,
  epsfig,
  xcolor,
  url,
  setspace,
  graphics,
  lipsum,
  lineno,
  placeins,
  hyperref
}

\makeatletter
\def\section{\@startsection{section}{1}%
      \z@{.7\linespacing\@plus\linespacing}{.5\linespacing}%
      {\normalfont\Large\bfseries\centering }}
\def\sectionL{\@startsection{section}{1}%
      \z@{.7\linespacing\@plus\linespacing}{.5\linespacing}%
      {\normalfont\Large\bfseries}}% NEW
\makeatother

\makeatletter
\def\paragraph{\@startsection{paragraph}{4}%
  \z@\z@{-\fontdimen2\font}%
  {\sffamily \bfseries }}
\makeatother

\newcommand{\bburl}[1]{\textcolor{blue}{\url{#1}}}

\definecolor{maroon}{rgb}{0.5, 0.0, 0.0}
\hypersetup{breaklinks=true,
            bookmarks=true,
            pdfauthor={Apoorva Lal},
             pdfkeywords = {},
            colorlinks=true,
            citecolor=maroon,
            urlcolor=blue,
            linkcolor=blue,
            pdfborder={0 0 0}}
\usepackage{float}
\usepackage{pdflscape}
\newcommand{\bland}{\begin{landscape}}
\newcommand{\eland}{\end{landscape}}

\numberwithin{equation}{section}

\newenvironment{itemize*}%
  {\begin{itemize}%
    \setlength{\itemsep}{0pt}%
    \setlength{\parskip}{0pt}}%
  {\end{itemize}}
\newenvironment{enumerate*}%
  {\begin{enumerate}%
    \setlength{\itemsep}{0pt}%
    \setlength{\parskip}{0pt}}%
  {\end{enumerate}}

\usepackage{
  amscd,
  amsfonts,
  amsmath,
  amssymb,
  amsbsy,
  bm, %
  dsfont,
  cancel, %
  latexsym,
  mathtools,
  graphicx,
  xcolor,
  xargs
}

\usepackage[colorinlistoftodos,prependcaption,]{todonotes}

\newcommand{\beq}{\begin{equation}}
\newcommand{\eeq}{\end{equation}}

\newcommand*\Bigpar[1]{\left( #1 \right )}

\newcommand{\ba}{\begin{array}}
\newcommand{\ea}{\end{array}}
\newcommand{\be}{\begin{enumerate}}
\newcommand{\ee}{\end{enumerate}}
\newcommand{\bi}{\begin{itemize}}
\newcommand{\ei}{\end{itemize}}
\newcommand{\bs}{\begin{align}\begin{split}\nonumber}
\newcommand{\bsnumber}{\begin{align}\begin{split}}
\newcommand{\es}{\end{split}\end{align}}

\newcommandx{\deriv}[2][1=x,2=f]{\nabla \, #2 \Bigpar{ #1 } }

\newcommand{\sumin}{\ensuremath{\sum_{i=1}^N}}

\newcommand{\inv}[1]{ \left({#1} \right)^{-1}}

\newcommand\frakfamily{\usefont{U}{yfrak}{m}{n}}
\DeclareTextFontCommand{\textfrak}{\frakfamily}

\newcommand{\ooN}{\frac{1}{N}}  %
\newcommand{\defeq}{\vcentcolon=}

\def\mbi#1{\boldsymbol{#1}} %
\def\ve#1{\mbi{#1}} %

\newcommand*{\Mat}[1]{\mathbf{#1}}

\newcommand{\wh}[1]{\widehat{#1}} %
\newcommand{\wt}[1]{\widetilde{#1}} %
\newcommand*\Ol[1]{\overline{#1}}

\newcommand{\E}{\mathbb{E}} %
\def\plim{\text{plim} \;}

\newcommand\indep{\protect\mathpalette{\protect\independenT}{\perp}}
\def\independenT#1#2{\mathrel{\rlap{$#1#2$}\mkern5mu{#1#2}}}

\newcommand{\Var}[1]{\mathbb{V}\left[#1\right]}
\newcommand{\Covar}[1]{\text{Cov}(#1)}
\newcommand{\hyp}[2]{
\ensuremath{H_0:#1 \ifhmode\quad\text{versus}\quad\fi\text{ vs. } H_1:#2}}

\newcommandx{\uniff}[1][1={a,b}]{\textrm{Unif}\left({#1}\right)}
\newcommandx{\unifd}[1][1={a,\ldots,b}]{\textrm{Unif}\left\{{#1}\right\}}
\newcommandx{\dunif}[3][1=x,2=a,3=b]{\frac{I(#2<#1<#3)}{#3-#2}}
\newcommandx{\dunifd}[3][1=x,2=a,3=b]{\frac{I(#2\le#1\le#3)}{#3-#2+1}}
\newcommandx{\punif}[3][1=x,2=a,3=b]{
\begin{cases} 0 & #1 < #2 \\ \frac{#1-#2}{#3-#2} & #2 < #1 < #3 \\ 1 & #1 > #3\\\end{cases}}
\newcommandx{\punifd}[3][1=x,2=a,3=b]{
\begin{cases} 0 & #1 < #2\\ \frac{\lfloor#1\rfloor-#2+1}{#3-#2} & #2 \le #1 \le #3 \\ 1 & #1 > #3\\ \end{cases}}

\newcommandx\bern[1][1=p]{\textrm{Bern}\left({#1}\right)}
\newcommandx\dbern[2][1=x,2=p]{#2^{#1} \left(1-#2\right)^{1-#1}}
\newcommandx\pbern[2][1=x,2=p]{\left(1-#2\right)^{1-#1}}

\newcommandx\bin[1][1={n,p}]{\textrm{Bin}\left(#1\right)}
\newcommandx\dbin[3][1=x,2=n,3=p]{\binom{#2}{#1}#3^#1\left(1-#3\right)^{#2-#1}}

\newcommandx\mult[1][1={n,p}]{\textrm{Mult}\left(#1\right)}
\newcommandx\dmult[3][1=x,2=n,3=p]{\frac{#2!}{#1_1!\ldots#1_k!}#3_1^{#1_1}\cdots#3_k^{#1_k}}

\newcommandx\hyper[1][1={N,m,n}]{\textrm{Hyp}\left({#1}\right)}
\newcommandx\dhyper[4][1=x,2=N,3=m,4=n]{\frac{\binom{#3}{#1}\binom{#2-#3}{#4-#1}}{\binom{#2}{#4}}}

\newcommandx\nbin[1][1={r,p}]{\textrm{NBin}\left({#1}\right)}
\newcommandx\dnbin[3][1=x,2=r,3=p]{\binom{#1+#2-1}{#2-1}#3^#2(1-#3)^#1}
\newcommandx\pnbin[3][1=x,2=r,3=p]{I_#3(#2,#1+1)}

\newcommandx\geo[1][1=p]{\textrm{Geo}\left(#1\right)}
\newcommandx\dgeo[2][1=x,2=p]{#2(1-#2)^{#1-1}}
\newcommandx\pgeo[2][1=x,2=p]{1-(1-#2)^#1}

\newcommandx\pois[1][1=\lambda]{\textrm{Po}\left({#1}\right)}
\newcommandx\dpois[2][1=x,2=\lambda]{\frac{#2^#1 e^{-#2}}{#1!}}
\newcommandx\ppois[2][1=x,2=\lambda]{e^{-#2}\sum_{i=0}^#1\frac{#2^i}{i!}}

\newcommandx\normall[1][1={\mu,\sigma^2}]{\mathcal{N}\left({#1}\right)}
\newcommandx\dnormall[3][1=x,2=\mu,3=\sigma]%
  {\frac{1}{#3\sqrt{2\pi}}\exp \Bigpar{-\frac{\left(#1-#2\right)^2}{2 #3^2}}}
\newcommandx\pnormall[1][1=x]{\Phi\left({#1}\right)}
\newcommandx\qnormall[1]{\Phi^{-1}\left({#1}\right)}

\newcommandx\mvn[1][1={\mu,\Sigma}]{\mathrm{MVN}\left({#1}\right)}

\newcommandx\ex[1][1=\lambda]{\textrm{Exp}\left(#1\right)}
\newcommandx\dex[2][1=x,2=\lambda]{#2e^{-#1 #2}}
\newcommandx\pex[2][1=x,2=\lambda]{1-e^{-#1 #2}}

\newcommandx\gam[1][1={\alpha,\lambda}]{\textrm{Gamma}\left({#1}\right)}
\newcommandx\dgamma[3][1=x,2=\alpha,3=\lambda]%
  {\frac{#3^{#2}}{\Gamma\left( #2 \right)} #1^{#2-1}e^{-#3#1}}

\newcommandx\invgamma[1][1={\alpha,\beta}]{\textrm{InvGamma}\left({#1}\right)}
\newcommandx\dinvgamma[3][1=x,2=\alpha,3=\beta]%
{\frac{#3^{#2}}{\Gamma\left(#2\right)}#1^{-#2-1}e^{-#3/#1}}
\newcommandx\pinvgamma[3][1=x,2=\alpha,3=\beta]%
{\frac{\Gamma\left(#2,\frac{#3}{#1}\right)}{\Gamma\left(#2\right)}}

\newcommandx\bet[1][1={\alpha,\beta}]{\textrm{Beta}\left(#1\right)}
\newcommandx\dbeta[3][1=x,2=\alpha,3=\beta]
{\frac{\Gamma\left(#2+#3\right)}{\Gamma\left(#2\right)\Gamma\left(#3\right)}#1^{#2-1}\left(1-#1\right)^{#3-1}}

\newcommandx\dir[1][1={\alpha}]{\textrm{Dir}\left(#1\right)}
\newcommandx\ddir[3][1=x,2=\alpha]{\frac{\Gamma\left(\sum_{i=1}^k #2_i\right)}{\prod_{i=1}^k\Gamma\left(#2_i\right)}\prod_{i=1}^k #1_i^{#2_i-1}}

\newcommandx\weibull[1][1={\alpha}]{\textrm{Dir}\left(#1\right)}
\newcommandx\dweibull[3][1=x,2=\lambda,3=k]{\frac{#3}{#2}
\left(\frac{#1}{#2}\right)^{#3-1} e^{-(#1/#2)^k}}

\newcommandx\chisq[1][1=k]{\chi_{#1}^2}

\newcommandx\zet[1][1=s]{\textrm{Zeta}\left(#1\right)}
\newcommandx\dzeta[2][1=x,2=s]{\frac{#1^{-#2}}{\zeta\left(#2\right)}}

\newtheorem{assum}{Assumption}

\usepackage{array}
\newcolumntype{L}[1]{>{\raggedright\let\newline\\\arraybackslash\hspace{0pt}}m{#1}}
\newcolumntype{C}[1]{>{\centering\let\newline\\\arraybackslash\hspace{0pt}}m{#1}}
\newcolumntype{R}[1]{>{\raggedleft\let\newline\\\arraybackslash\hspace{0pt}}m{#1}}

\usepackage[foot]{amsaddr}

\usepackage[
  hyperref=true,
  url=false,
  isbn=false,
  backref=true,
  maxcitenames=3,
  backend=biber,
  style=authoryear,
  citestyle=authoryear,
  natbib=true]{biblatex}

\renewbibmacro{in:}{}
\title[How much should we trust instrumental variables]{How Much Should We Trust Instrumental Variable Estimates in Political Science? Practical Advice Based on 67 Replicated Studies}
\author{Apoorva Lal}
\address{Independent Researcher}
\email{lal.apoorva@gmail.com}

\author{Mac Lockhart}
\address{Institution for Social and Policy Studies, Yale University, New Haven, CT 06511, USA.}
\email{mwlockha@ucsd.edu}

\author{Yiqing Xu}
\address{Department of Political Science, Stanford University, Stanford, CA 94305, USA.}
\email{yiqingxu@stanford.edu}

\author{Ziwen Zu}
\address{Department of Political Science, University of California, San Diego, La Jolla, CA 92093, USA.}
\email{zzu@ucsd.edu}

\begin{document}
\begin{abstract}
Instrumental variable (IV) strategies are widely used in political science to establish causal relationships, but the identifying assumptions required by an IV design are demanding, and assessing their validity remains challenging. In this paper, we replicate 67 articles published in three top political science journals from 2010-2022 and identify several concerning patterns. First, researchers often overestimate the strength of their instruments due to non-i.i.d. error structures such as clustering. Second, IV estimates are often highly uncertain, and the commonly used $t$-test for two-stage-least-squares (2SLS) estimates frequently underestimates the uncertainties. Third, in most replicated studies, 2SLS estimates are significantly larger than ordinary-least-squares estimates, and their ratio is inversely related to the strength of the instrument in observational studies---a pattern not observed in experimental ones---suggesting potential violations of unconfoundedness or the exclusion restriction in the former.  We provide a checklist and software to help researchers avoid these pitfalls and improve their practice.

\end{abstract}

% \keywords{instrumental variables, two-stage-least-squares, replications, weak instruments, unconfoundedness, exclusion restriction, publication bias, meta analysis}

\maketitle

\section{Introduction} \label{intro}

The instrumental variable (IV) approach is a widely used empirical method in the social sciences, including political science, for establishing causal relationships. It is often used when selection on observables is implausible, experimentation is infeasible or unethical, and rule-based assignments that allow for sharp regression-discontinuity (RD) designs are unavailable. In recent years, there has been a growing number of articles published in top political science journals, such as the \emph{American Political Science Review} (APSR),  \emph{American Journal of Political Science} (AJPS), and \emph{Journal of Politics} (JOP), that use IV as a primary causal identification strategy. This trend can be traced back to the publication of \emph{Mostly Harmless Econometrics} \parencite{angrist2008mostly}, which popularized the modern interpretation of IV designs, and \cite{sovey2011instrumental}, which clarifies the assumptions required by an IV design and provides a useful checklist for political scientists.

\begin{figure}[!ht]
\begin{minipage}{1\linewidth}{
\begin{center}
\includegraphics[width=0.9\textwidth]{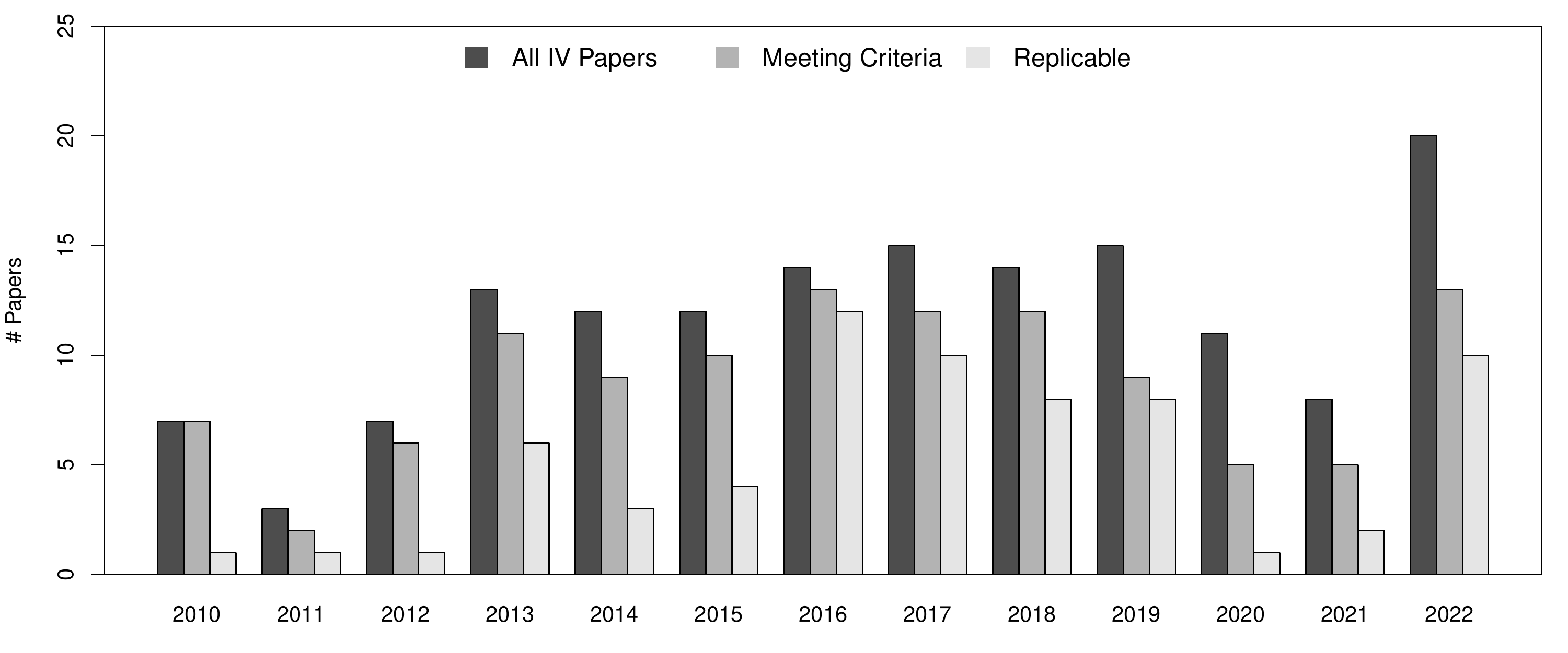}
\end{center}\vspace{-1em}
\caption{IV studies published in the \emph{APSR}, \emph{AJPS}, and \emph{JOP}. Our criteria rule out IV models appearing in the online appendix only, in dynamic panel settings, with multiple endogenous variables, and with nonlinear link functions. Non-replicability is primarily due to a lack of data and/or coding errors.}}\label{fig:pubs.year}
\end{minipage}
\end{figure}

Despite its popularity, the IV approach has faced scrutiny from researchers who note that two-stage least-squares (2SLS) estimates are often much larger in magnitude than ``naïve'' ordinary-least-squares (OLS) estimates, even when the main concern with the latter is upward omitted-variables bias.\footnote{For example, in the 2016 National Bureau of Economic Research--Political Economy Meeting, following a presentation of a study using an IV approach, the late political economist Alberto Alesina asked the audience: ``How come 2SLS estimates are always five times bigger than OLS estimates in political economy?'' We dedicate this paper to him for his seminal contributions to the field of political economy.} Others have raised concerns about the validity of the commonly used inferential method for 2SLS estimation \parencite[e.g.][]{Lee2020-mi,Young2022}.

These observations motivate our systematic examination of the use of IVs in the empirical political science literature. We set out to replicate all studies published in the APSR, AJPS, and JOP during the past thirteen years (2010-2022) that use an IV design with a single endogenous variable as one of the main identification strategies.\footnote{Focusing on design with a single endogenous variable allows us to calculate the first-stage correlation coefficient (or $R^2$) and apply tools such as the Anderson-Rubin (AR) test and the $tF$ test (when there is only a single instrument). Moreover, we find it difficult to justify the exclusion restriction in a multiple-treatment-multiple-instrument setting in the first place.}  Out of 114 articles meeting this criterion, 71 have complete replication materials online, which is concerning in itself. We successfully replicated at least one primary IV result in 67 out of the remaining 71 articles. Among the 67 articles, three articles feature two distinct IV designs, each yielding two separate replicable IV results.

Using data from these 70 IV designs, we conduct a programmatic replication exercise and find three troubling patterns. First, a significant number of IV designs in political science overestimate the first-stage partial $F$-statistic by failing to adjust standard errors (SEs) for factors such as heteroskedasticity, serial correlation, or clustering structure. Using the effective $F$-statistic \parencite{Olea2013-pa}, we find that at least 11\% of the published IV studies rely on what econometricians call ``weak instruments,'' the consequences of which have been well-documented in the literature. See \cite{andrews2019weak} for a comprehensive review.

Second, obtaining valid statistical inferences for IV estimates remains challenging. Almost all studies we have replicated rely on $t$-tests for the 2SLS estimates based on analytic SEs and traditional critical values (such as 1.96 for statistical significance at the 5\% level). Using analytic SEs, IV estimates are already shown to be much more imprecise than OLS estimates. When employing bootstrapping procedures, the AR test, or the $tF$ procedure---an $F$-statistic-dependent $t$-test \parencite{Lee2020-mi}---for hypothesis testing, we find that 17-35\% of the designs cannot reject the null hypothesis of no effect at the 5\% level. In contrast, only 10\% of studies based on originally reported SEs or $p$-values fail to reject the null hypothesis. This discrepancy suggests that many studies may have underestimated the uncertainties associated with their 2SLS estimates.

What is even more concerning is that an IV approach can produce larger biases than OLS when weak instruments amplify biases due to failures of IVs' unconfoundedness or exclusion restrictions. We observe that in 68 out of the 70 designs (97\%), the 2SLS estimates have a larger magnitude than the naïve OLS estimates obtained from regressing the outcome on potentially endogenous treatment variables and covariates; 24 of these (34\%) are at least five times larger. This starkly contrasts with the common rationale for using IV, which is to mitigate the upward bias in treatment effect estimates from OLS. Moreover, we find the ratio between the magnitudes of the 2SLS and OLS estimates is strongly negatively correlated with the strength of the first stage among studies that use non-experimental instruments, and the relationship is almost nonexistent among experimental studies. While factors such as heterogeneous treatment effects and measurement error might be at play, we contend that this phenomenon primarily stems from a combination of weak instruments and the failure of unconfoundedness or the exclusion restriction. Intuitively, because the 2SLS estimator is a ratio, an inflated numerator from invalid instruments paired with a small denominator due to a weak first stage leads to a disproportionately large estimate. Publication bias and selective reporting exacerbate this issue.

What do these findings imply for IV studies in political science? First, the traditional $F$-tests for IV strength, particularly when using classic analytic SEs, often mask the presence of weak instruments. Second, when operating with these weak instruments, particularly in over-identified scenarios, traditional $t$-tests do not adequately represent the considerable uncertainty surrounding the 2SLS estimates, paving the way for selective reporting and publication bias. Last but not least, many 2SLS estimates likely bear significant biases due to violations of unconfoundedness or the exclusion restriction, and weak instruments further exacerbate these biases. While we cannot pinpoint exactly which estimates are problematic, these issues seem to be pervasive across observational IV studies. The objective of this paper, however, is not to discredit existing IV research or dissuade scholars from using IVs. On the contrary, our intent is to caution researchers against the pitfalls of ad-hoc justifications for IVs in observational research and provide constructive recommendations for future practices. These suggestions include accurately quantifying the strength of instruments, conducting valid inference for IV estimates, as well as implementing additional validation exercises, such as placebo tests, to bolster the identifying assumptions.

Our work builds on a growing literature evaluating IV strategies in social sciences and offering methods to improve empirical practice. Notable studies include \cite{Young2022}, which finds IV estimates to be more sensitive to outliers and conventional $t$-tests to understate uncertainties; \cite{jiang2017have}, which observes larger IV estimates in finance journals and attributed this to exclusion restriction violations and weak instruments; \cite{mellon2020rain}, which emphasizes the vulnerability of weather instruments; \cite{dieterle2016simple}, which develops a quadratic over-identification test and discovers significant non-linearities in the first stage regression; \cite{FeltonStewart2022}, which finds unstated assumptions and a lack of weak-instrument robust tests in top sociology journals; and \cite{cinelli2022omitted}, which proposes a sensitivity analysis for IV designs in an omitted variable bias framework. This study is the first comprehensive replication effort focusing on IV designs in political science and uses data to shed light on the consequences of weak instruments interacting with failures of unconfoundedness or the exclusion restriction.

\section{Theoretical Refresher} \label{theory}

In this section, we offer a brief overview of the IV approach, including the setup, the key assumptions, and the 2SLS estimator. We then discuss potential pitfalls and survey several inferential methods. To cover the vast majority of IV studies in political science, we adopt a traditional constant treatment effect approach, which imposes a set of parametric assumptions. For example, of our replication sample, 51 designs (73\%) employ continuous treatment variables, and 49 (70\%) use continuous IVs. Most of these studies make no reference to treatment effect heterogeneity and are ill-suited for the local average treatment effect framework \parencite{angrist1996identification}.

Apart from the canonical use of IVs in addressing non-compliance in experimental studies, we observe that in the majority of the articles we review, researchers use IVs to establish causality between a single treatment variable $d$ and an outcome variable $y$ in observational settings. The basic idea of this approach is to use a vector of instruments $z$ to isolate ``exogenous'' variation in $d$ (i.e., the variation in $d$ that is not related to potential confounders) and estimate its causal effect on $y$. For simplicity, we choose not to include any additional exogenous covariates in our discussion. This is without loss of generality because, by the Frisch-Waugh-Lovell theorem, we can remove these variables by performing a regression of $y$, $d$, and each component of $z$ on the controls and then proceeding with our analysis using the residuals instead.

\subsection{Identification and Estimation}

Imposing a set of parametric assumptions, we define a system of simultaneous equations:
\begin{align}
\text{Structural equation:}\quad y =&\ \tau_{0} + \tau d  + \varepsilon\label{eq1} \\
\text{First-stage equation:}\quad   d =&\ \pi_0 + \pi' z  + \nu\label{eq2}
\end{align}
in which $y$ is the outcome variable; $d$ is a scalar treatment
variable; $z$ is a vector of instruments for $d$; and $\tau$ captures the (constant)
treatment effect and is the key quantity of interest. The error terms $\varepsilon$ and $\nu$ may be correlated. The endogeneity problem for $\tau$ in Equation~(\ref{eq1}) arises when $d$ and $\varepsilon$ are correlated, which renders $\hat\tau_{OLS}$ from a naïve OLS regression of $y$ on $d$ inconsistent. This may be due to several reasons: (1) unmeasured omitted variables correlated with both $y$ and $d$; (2) measurement error in $d$, or (3) simultaneity or reverse causality, which means $y$ may also affect $d$.  Substituting $d$ in Equation~(\ref{eq1}) using Equation~(\ref{eq2}), we have the reduced form equation:
\begin{equation}
\text{Reduced form:}\quad y =
\underbrace{(\alpha + \tau \pi_0)}_{\gamma_0} +
\underbrace{(\tau \pi)^{\prime}}_{\gamma'} z + (\tau\nu + \varepsilon).
\end{equation}
Substitution establishes that $\gamma = \tau \pi$, rearranging
yields $\tau = \frac{\gamma}{\pi}$  (assuming a single instrument, but the intuition carries over to cases with multiple instruments). The IV estimate, therefore, is the ratio of the reduced-form and first-stage
coefficients. To identify $\tau$, we make the following assumptions
\parencite[Chapter 12]{greene2003econometric}.

\begin{assum}[Relevance] $\pi \neq 0$. \label{assumption:rel}
This assumption requires that the IVs can predict the treatment variable, and is therefore equivalently stated as $d \not \indep z$.
\end{assum}

\begin{assum}[Exogeneity: unconfoundedness \& the exclusion restriction]\label{assumption:exog} $\E[\varepsilon] =0$ and\\ $\Covar{z, \varepsilon} 
= 0$.  This assumption is satisfied when unconfoundedness (random or quasi-random assignment of $z$) and the exclusion restriction (no direct effect of $z$ on $y$ beyond $d$) are met. 
\end{assum}
Conceptually, unconfoundedness and the exclusion restriction are two distinct assumptions and should be justified separately in a research design. However, because violations of either assumption lead to the failure of the 2SLS moment condition, $\E[z\varepsilon] = 0$, and produce observationally equivalent outcomes, we consider both to be integral components of Assumption~2.

Under Assumptions~\ref{assumption:rel} and \ref{assumption:exog}, the 2SLS estimator is shown to be consistent for the structural parameter $\tau$. Consider a sample of $N$ observations. We can write $\Mat{d} = (d_1, d_2, \cdots, d_N)'$ and $\Mat{y}= (y_1, y_2, \cdots, y_N)'$ as $(N\times 1)$ vectors of the treatment and outcome data, and $\Mat{z}= (z_1, z_2, \cdots, z_N)'$ as an $(N\times p_z)$ matrix of instruments in which $p_z$ is the number of instruments. The 2SLS estimator is written as follows: 
\begin{equation} \hat\tau_{\text{2SLS}} = \inv{\Mat{d}'
\Mat{P}_{z} \Mat{d}}  \Mat{d}' \Mat{P}_{z} \ve{y}
\end{equation}
in which $\Mat{P}_{z} = \Mat{z}\inv{\Mat{z}' \Mat{z}} \Mat{z}'$ is the hat-maker matrix from the first stage which projects the endogenous treatment variable $\Mat{d}$ into the column space of $\Mat{z}$, thereby in expectation preserving only the exogenous variation in $\Mat{d}$ that is uncorrelated with $\varepsilon$. This formula permits the use of multiple instruments, in which case the model is said to be ``overidentified.'' The 2SLS estimator belongs to a class of generalized method of moments (GMM) estimators taking advantage of the moment condition $\E[z\varepsilon] =0$, including the two-step GMM \parencite{hansen1982large} and limited information maximum likelihood (LIML) estimators  \parencite{anderson1982evaluation}. We use the 2SLS estimator throughout the replication exercise because of its simplicity and because every single paper in our replication sample uses it in at least one specification.

When the model is exactly identified, i.e., the number of treatment variables equals the number of instruments, the 2SLS estimator can be simplified as the IV estimator: $\hat{\tau}_{\text{2SLS}} = \hat{\tau}_{\text{IV}} = \inv{\Mat{z}'\Mat{d}} \Mat{z}'\Mat{y}$. In the case of one instrument and one treatment, the 2SLS estimator can also be written as a ratio of two sample covariances: $\hat{\tau}_{2SLS} = \hat{\tau}_{IV}
= \frac{\hat\gamma}{\hat\pi} = \frac{\widehat{\rm Cov}(y, z)}{\widehat{\rm Cov}(d, z)}$, which illustrates that the 2SLS estimator is a ratio between reduced-form and first-stage coefficients in this special case. This further simplifies to a ratio of the differences in means when $z$ is binary, which is
called a Wald estimator.

\subsection{Potential Pitfalls in Implementing an IV Strategy} %
\label{sub:problems_in_iv_estimation}

The challenges with 2SLS estimation and inference are mostly due to violations of Assumptions~\ref{assumption:rel} and \ref{assumption:exog}. Such violations can result in (1) significant uncertainties around 2SLS estimates and size distortion for $t$-tests due to weak instruments even when the exogeneity assumption is satisfied; and (2) potentially larger biases in 2SLS estimates compared to OLS estimates when both assumptions are violated.

\paragraph*{Inferential problem due to weak instruments.} Since the IV coefficient is a ratio, the weak instrument problem is a ``divide-by-zero'' problem, which arises when $\Covar{z, d} \approx 0$ (i.e., when the relevance assumption is violated). The instability of ratio estimators like $\wh\tau_{\text{2SLS}}$ when the denominator is approximately zero has been extensively studied going back to \cite{fieller1954some}. The conventional wisdom in the past two decades has been that the first-stage partial $F$-statistic needs to be bigger than 10, and it should be clearly reported \parencite{Staiger1997-lo}. The cutoff, as a rule of thumb, is chosen based on simulation results to meet two criteria under i.i.d. errors: (1) in the \emph{worst} case, the bias of the 2SLS estimator does not exceed 10\% of the bias of the OLS estimator, and (2) a $t$-test based on the 2SLS estimator with a size of 5\% does not lead to size over 15\%. These problems are further exacerbated in settings where units belong to clusters with strong within-cluster correlation, where a small number of observations or clusters may heavily influence estimated results \parencite{Young2022}. Recently, however, \cite{angrist2023one} argue that the conventional inference strategies are reliable in just-identified settings with independent errors. The weak instrument issue is indeed most concerning in heavily over-identified scenarios.

The literature has discussed at least three issues caused by weak instruments when the exogeneity assumption is satisfied. First, under i.i.d. errors, a weak first stage exacerbates the finite-sample bias of the 2SLS estimator toward the inconsistent OLS estimator, thereby reproducing the endogeneity problem that an IV design was meant to solve \parencite{Staiger1997-lo}. Additionally, when the first stage is weak, the 2SLS estimator may not have a mean; its median is centered around the OLS coefficient \parencite{hirano2015location}. Second, the 2SLS estimates become very imprecise. To illustrate, a commonly used variance estimator for
$\hat{\tau}_{IV}$ is $\hat{\mathbb{V}}(\hat{\tau}_{IV}) \approx \hat{\sigma}^2/(\sumin (d_i - \Ol{d})^2 R_{dz}^2) = \hat{\mathbb{V}}(\hat{\tau}_{OLS}) /R_{dz}^2$, in which $\hat{\sigma}^2$ is a variance estimator for the error term and
$R^2_{dz}$ is the first stage $R^2$. $\hat{\mathbb{V}}(\hat{\tau}_{IV})$
is generally larger than $\hat{\mathbb{V}}(\hat{\tau}_{OLS})$ and increasing in $1/R^2_{dz}$. A third and related issue is that the $t$-tests are of the wrong size and the $t$-statistics do not follow a $t$-distribution \parencite{nelson1990some}. This is because the distribution of $\hat{\tau}$ is derived from its linear approximation of $\hat{\tau}$ in ($\hat{\gamma}, \hat{\pi}$), wherein normality of the two OLS coefficients implies the normality of their ratio. However, this normal approximation breaks down when $\hat{\pi} \approx 0$. Moreover, this approximation failure cannot generally be rectified by bootstrapping \parencite{andrews2009validity}; \cite{Young2022} argues that it nevertheless allows for improved inference when outliers are present. Overall, valid IV inference relies crucially on strong IVs.

Generally, there are two approaches to conducting inference in an IV design: pretesting and direct testing. The pretesting approach involves using an $F$-statistic to test the first stage strength, and if it exceeds a certain threshold (e.g., $F > 10$), proceeding to test the null hypothesis about the treatment effect (e.g., $\tau = 0$). Nearly all reviewed studies employ this approach. The direct testing approach, in contrast, does not rely on passing a pretest. We examine four inferential methods for IV designs, with the first three related to pretesting and the last one being a direct test.

First, \cite{Olea2013-pa} propose the effective $F$-statistic for both just-identified and over-identified settings and accommodates robust or cluster-robust SEs. The effective $F$ is a scaled version of the first-stage $F$-statistic and is computed as $F_{\text{Eff}} = \hat{\pi}'\hat{Q}_{\text{ZZ}} \hat{\pi} / \text{tr}(\hat{\Sigma}_{\pi \pi} \hat{Q}_{\text{ZZ}})$, where $\hat{\Sigma}_{\pi \pi}$ is the variance-covariance matrix of the first stage regression, and $\hat{Q}_{\text{ZZ}} = \ooN \sumin z_i z_i'$. In just-identified cases, $F_{\text{Eff}}$ is the same as an $F$-statistic based on robust or cluster-robust SEs. The authors derive the critical values for $F_{\text{Eff}}$ and note that the statistic and corresponding critical values are identical to the better-known robust $F$-statistic $\hat{\pi} \hat{\Sigma}_{\pi \pi}^{-1} \hat{\pi}$ and corresponding \cite{stock2005asymptotic} critical values.  $F_{\text{Eff}}>10$ is shown to be a reasonable rule of thumb under heteroskedasticity in simulations \parencite{Olea2013-pa, andrews2019weak}.

Second, \cite{Young2022} recommends researchers report two types of bootstrap confidence intervals (CIs), \emph{bootstrap-c} and \emph{bootstrap-t}, for $\hat{\tau}_{2SLS}$ under non-i.i.d. errors with outliers, which is common in social science settings. They involve $B$ replications of the following procedure: (1) sample $n$ triplets $(y_i^*, d_i^*, z_i^*)$ independently and with replacement from the original sample (with appropriate modifications for clustered dependence) and (2) on each replication, compute the 2SLS coefficient and SE, as well as the corresponding test statistic $t^* = \hat{\tau}^*_{\text{2SLS}} / \hat{\text{SE}} (\hat{\tau}^*_{\text{2SLS}})$. The \emph{bootstrap-c} method calculates the CIs by taking the $\alpha/2$ and $(1-\alpha/2)$ percentiles of the bootstrapped 2SLS coefficients $\hat{\tau}_{\text{2SLS}}^*$, while the \emph{bootstrap-t} method calculates the percentile-$t$ refined CIs by plugging in the $\alpha/2$ and $(1-\alpha/2)$ percentile of the bootstrapped $t$ statistics into the expression $\hat{\tau}_{\text{2SLS}} \pm t^*_{\alpha \mid 1 - \alpha} \hat{\text{SE}}(\hat{\tau}^*_{\text{2SLS}})$. \cite{hall1996bootstrap} show that \emph{bootstrap-t} achieves an asymptotic refinement over \emph{bootstrap-c}. Note that $t$-tests based on bootstrapped SEs may be overly conservative \parencite{Hahn2021-lq} and, hence, are not recommended.

Third, in just-identified single treatment settings, \cite{Lee2020-mi} propose the $tF$ procedure that smoothly adjusts the $t-$ratio inference based on the first-stage $F$-statistic, which improves upon the ad-hoc screening rule of $F > 10$. The adjustment factor applied to 2SLS SEs is based on the first stage $t-$ratio $\hat{f} \defeq \hat{\pi}/\sqrt{\hat{\mathbb{V}}(\hat{\pi})}$, with the first stage $\hat{F} = \hat{f}^2$, and relies on the fact that the distortion from employing the standard 2SLS $t$-ratio $\hat{t} \defeq \hat{\tau}/\sqrt{\hat{\mathbb{V}}(\hat{\tau})}$ can be quantified in terms of an $\hat{F}$ statistic, which gives rise to a set of critical values for a given pair of $\hat{t}$ and $\hat{F}$. The authors also show that if no adjustment is made to the $t$-test's critical value (e.g., using 1.96 as the threshold for 5\% statistical significance), a first stage $\hat{F}$ of 104.7 is required to guarantee a correct size of $5\%$ for a two-sided $t$-test for the 2SLS coefficient.

Finally, where there is one endogenous treatment variable, the AR procedure, which is essentially an $F$-test on the reduced form, is a direct inferential method robust to weak instruments \parencite{anderson1949estimation, chernozhukov2008reduced}. Without loss of generality, assume that we are interested in testing the null hypothesis that $\tau = 0$, which then implies that the reduced form coefficient from regressing $y$ on $z$ is zero, i.e., $\gamma = 0$. This motivates the following procedure: given a set $\mathcal{T}$ of potential values for $\wt{\tau}$, for each value $\wt{\tau}$, construct $\wt{y} = y - d \wt{\tau}$, and regress $\wt{y}$ on $z$ to obtain a point estimate $\wt{\gamma}$ and (robust, or cluster robust) covariance matrix $\wt{\mathbb{V}}(\wt{\gamma})$, and construct a Wald statistic $\wt{W}_s(\wt{\gamma}) \defeq \wt{\gamma}' \wt{\mathbb{V}}(\wt{\gamma})^{-1} \wt{\gamma}$. Then, the AR CI (or confidence set) is the set of $\wt{\gamma}$ such that $\wt{W}_s(\wt{\gamma}) \leq c(1-p)$ where $c(1-p)$ is the $(1-p)^{\text{th}}$ percentile of the $\chi^2_1$ distribution. The AR test requires no pretesting and is shown to be the uniformly most powerful unbiased test in the just-identified case \parencite{moreira2009tests}. It is less commonly used than pretesting procedures possibly because researchers are more accustomed to using $t$-tests than $F$-tests and reporting SEs rather than CIs. A potential limitation of the AR test is that its CIs can sometimes be empty or disconnected, and therefore lack a Bayesian interpretation under uninformative priors.\footnote{We thank Guido Imbens for highlighting this point.}

\paragraph*{Bias amplification and the failure of the exogeneity assumption.} When the number of instruments is bigger than the number of endogenous treatments, researchers can use an over-identification test to gauge the plausibility of Assumption~\ref{assumption:exog}, the exogeneity assumption \parencite{arellano2002sargan}. However, such a test is often underpowered and has bad finite sample properties \parencite{davidson2015bootstrap}. In just-identified cases, Assumption~\ref{assumption:exog} is not directly testable. When combined with weak instruments, even small violations of unconfoundedness or the exclusion restriction can produce inconsistency. This is because: $\plim\hat{\tau}_{IV} = \tau + \frac{\Covar{z, \varepsilon}}{\Covar{z, d}}$. When $\Covar{z, d} \approx 0$, even small violations of exogeneity, i.e., $\Covar{z, \varepsilon} \neq 0$, will enlarge the second term, resulting in large biases. Thus, the two identifying assumption failures exacerbate each other: having weak instruments compounds problems from confounding or exclusion restriction violations, and vice versa. With invalid instruments, it is likely that the asymptotic bias of the 2SLS estimator is much greater than that of the OLS estimator, i.e., $\left|\frac{\Covar{z, \varepsilon}}{\Covar{z, d}}\right| \gg  \left|\frac{\Covar{d, \varepsilon}}{\Var{ d}}\right|$ in the single instrument case.\footnote{We are not the first to make this argument. According to \cite[][p. 34]{hahn2005estimation}: ``[T]he empirical finding that the 2SLS  estimate increases compared to the OLS estimate may indicate that the instrument is not orthogonal to the stochastic disturbance.  The resulting bias can be substantial. Indeed, it could exceed the OLS bias,  leading to an increase in the estimated 2SLS coefficient over the estimated OLS coefficient.''}

While the inferential problem can be alleviated by employing alternative inferential methods as described above, addressing violations of unconfoundedness or the exclusion restriction is more challenging since it is fundamentally a research design issue that should be tackled at the design stage. Researchers often devote significant effort to arguing for unconfoundedness and exclusion restrictions in their settings. In Section A3 of the SM, we provide an exposition of the zero-first-stage (ZFS) test \parencite{bound2000compulsory}, which is essentially a placebo test on a subsample where the instrument is expected to be uncorrelated with the treatment, to help researchers gauge the validity of their instruments. These estimates can then be used to debias the 2SLS estimate using the methods proposed in \cite{Conley2012-mu}.

\section{Data and Types of Instruments} \label{desc}

In this section, we first discuss our case selection criteria and replication sample, which is the focus of our subsequent analysis. We then describe the types of instruments in the replicable studies.

\paragraph*{Data.} We examine all empirical articles published in the
APSR, AJPS, and JOP from 2010 to 2022 and
identify studies that use an IV strategy as one of the main
identification strategies, including articles that use binary or
continuous treatments and that use a single or multiple instruments.
We use the following criteria: (1) the discussion of the
IV result needs to appear in the main text and support a main argument
in the paper; (2) we consider linear models only; in other words, articles that use discrete outcome models are excluded from our sample;%
\footnote{We expect the issue identified in this paper to be present, if not more pronounced, with nonlinear IVs. With nonlinear IVs, weak instruments correspond to weak identification in GMM estimations for some or all unknown parameters. Consequently, weak identification results in non-normal distributions even in large samples, rendering conventional IV or GMM inferences unreliable \parencite{stock2002survey}.}
(3) we exclude articles that include multiple endogenous variables in a single specification (multiple endogenous variables in separate specifications are included); (4) we exclude articles that use IV or GMM estimators in a dynamic panel setting because the validity of the instruments (for example, $y_{t-2}$ affects $y_{t-1}$ but not $y_{t}$) is often not grounded in theories or substantive knowledge; these applications are subject to a separate set of empirical issues, and their poor performance has been discussed in the literature \parencite[e.g.,][]{bun2010weak}. These criteria result in 30 articles in the APSR, 33 articles in the AJPS, and 51 articles in the JOP. We then strive to find replication materials for these articles from public data-sharing platforms, such as the Harvard Dataverse, and the
authors' websites. We are able to locate complete replication materials for 76 (62\%) articles. However, code completeness and documentation quality vary widely. Since 2016-2017, data availability has significantly improved, thanks to new editorial policies that require authors to make replication materials publicly accessible \parencite{key2016we}. Starting in mid-2016 for AJPS and early-2021 for JOP, both journals introduced a policy requiring third-party verification of full replicability as a prerequisite for publication, although not all data are made public. We view these measures as significant advancements.

\begin{table}[htbp]
  \centering\small
  \caption{Data availability and replicability of IV articles.}
    \label{tb:sample}%
    \begin{tabular}{lccccc}\hline\hline
          \multicolumn{1}{p{5em}}{} & \multicolumn{1}{p{5em}}{} &
          \multicolumn{1}{p{5em}}{\centering Incomplete} & \multicolumn{1}{p{5em}}{\centering Incomplete} & \multicolumn{1}{p{5em}}{\centering Replication} & \multicolumn{1}{p{5em}}{}  \\
            & \#All Articles  & Data & Code & Error  & Replicable  \\ \hline
    APSR  & 33  & 16  & 0 & 3 & 14 (42\%) \\
    AJPS  & 30  & 3  & 1 & 1 & 25 (83\%) \\
    JOP   & 51  & 19 & 3  & 1 & 28 (55\%)\\ \hline
    Total & 114 & 38 & 4 & 5 & 67 (59\%) \\\hline
    \end{tabular}%
\end{table}%

Using data and code from the replication materials, we set out to replicate the main IV results in these 76 articles with complete data. Our replicability criterion is simple: As long as we can exactly replicate \emph{one} 2SLS point estimate that appears in the paper, we deem the paper replicable. We do not aim at exactly replicating SEs, $z$-scores, or level of statistical significance for the 2SLS estimates because they involve the choice of the inferential method. After much effort and hundreds of hours of work, we are able to
replicate the main results of 67 articles.\footnote{For three articles, we
are able to produce the 2SLS estimates with perfectly executable code;
however, our replicated estimates are inconsistent with what was
reported in the original studies. We suspect the inconsistencies are
caused by data rescaling or misreporting; hence, we keep them in the
sample.} The low replication rate is consistent with what is reported in \cite{Hainmueller2019-wx} and \cite{chiu2023what}. The main reasons for failures of replication are incomplete data (38 articles), incomplete code or poor documentation (4 articles), and replication errors (5 articles). Table~\ref{tb:sample} presents summary statistics on data availability and replicability of IV articles for each of the three journals. The rest of this paper focuses on results based on these 67 replicable articles (and 70 IV designs).

\paragraph*{Types of instruments.} Inspired by \cite{sovey2011instrumental}, in Table~\ref{tb:iv.type}, we summarize the types of IVs in the replicable designs, although our categories differ from theirs to reflect changes in the types of instruments used in the discipline. These categories are ordered based on the strength of the design, in our view, for an IV study.

The first category is randomized experiments. These articles employ randomization, designed and conducted by researchers or a third party, and use 2SLS estimation to tackle non-compliance. With random assignment, our confidence in the exogeneity assumption increases because unconfoundedness is guaranteed by design and the direct effect of the instrument on the outcome is easier to rule out than without random assignment. For instance, \cite{alt2016} use assignment to an information treatment as an instrument for economic beliefs to understand the relationship between economic expectations and vote choice. Compared to IV articles published before 2010, the proportion of articles using experiment-generated IVs has increased significantly (from 2.9\% to 17.1\%) due to the growing popularity of experiments.

\begin{table}[htbp]
  \centering
  \caption{Types of Instruments} \label{tb:iv.type}%
    \begin{tabular}{lcc}\hline\hline
    Type & \multicolumn{1}{c}{\#Articles} & \multicolumn{1}{c}{Percentage \%} \\\hline
    \textbf{Experiment} & 12     & 17.1 \\ 
    \textbf{Rules \&  policy changes}  & 7     & 10.0 \\
    $\qquad$ Fuzzy RD & $\qquad$ 4     & $\qquad$ 5.7 \\ 
    $\qquad$ Change in exposure & $\qquad$ 3     & $\qquad$ 4.3 \\
    \textbf{Theory}    & 42    & 60.0 \\
    $\qquad$ Weather/climate/Geography & $\qquad$ 13     & $\qquad$ 18.6 \\
    $\qquad$ Treatment diffusion & $\qquad$ 2    & $\qquad$ 2.9 \\
    $\qquad$ History & $\qquad$ 11    & $\qquad$ 15.7 \\
    $\qquad$ Others & $\qquad$ 16 & $\qquad$ 22.9 \\
    \textbf{Econometrics} & 9     & 12.9 \\
    $\qquad$ Interactions/``Bartik'' & $\qquad$ 7     & $\qquad$ 10.0 \\
    $\qquad$ Lagged treatment & $\qquad$ 1     & $\qquad$ 1.4 \\
    $\qquad$ Empirical test & $\qquad$ 1     & $\qquad$ 1.4 \\\hline
    \textbf{Total} & 70    & 100.0 \\\hline
    \end{tabular}%
\end{table}%

Another category consists of instruments derived from explicit rules on observed covariates, creating quasi-random variations in the treatment. \cite{sovey2011instrumental} refer to this category as ``Natural Experiment.'' We avoid this terminology because it is widely misused and limit this category to two circumstances: fuzzy RD designs and variation in exposure to policies due to time of birth or eligibility. For example, \cite{kim2019} leverages a reform in Sweden that requires municipalities above a population threshold to adopt direct democratic institutions. \cite{dinas2014} uses eligibility to vote based on age at the time of an election as an instrument for whether respondents did vote. While rule-based IVs offer a pathway to credible causal inference, recent studies have raised concerns about their implementation, highlighting issues of insufficient power in many RD designs \parencite{stommes2023reliability}.

The next category is ``Theory,'' where the authors justify unconfoundedness and the exclusion restriction using social science theories or substantive knowledge. Over a decade after \cite{sovey2011instrumental}'s survey, it remains the most prevalent category among IV studies in political science, at around 60\%. We divide theory-based IVs into four subcategories: geography/climate/weather, treatment diffusion, history, and others. First, Many studies in the theory category justify the choices of their instruments based on geography, climate, or weather conditions. For example, \cite{zhu2017} uses weighted geographic closeness as an instrument for the activities of multinational corporations; \cite{hager2019} use mean elevation and distance to rivers to instrument equitable inheritance customs; \cite{henderson2016mediating} use rainfall around Election Day as an instrument for Democratic vote margins. Relatedly, several studies base their choices on regional diffusion of treatment. For example, \cite{dube2015} use US military aid to countries outside Latin America as an instrument for US military aid to Colombia. \cite{dorsch_maarek2019} use the regional share of democracies as an instrument for democratization in a country-year panel.%
\footnote{While authors often argue that weather or geography is quasi-randomly imposed, it is typically harder to claim they only affect the outcome through the treatment variable. For example, \cite{mellon2020rain} contends that, while instruments like rain may be quasi-random, researchers have pinpointed several mechanisms through which it influences key political outcomes. \cite{betz2018use} argue that spatial instruments are rarely valid because of cross-sectional interdependence and simultaneity. Inference also presents challenges.}
Third, historical instruments derive from past differences between units unrelated to current treatment levels. For example, \cite{vernby2013} uses historical immigration levels as an instrument for the current number of non-citizen residents. Finally, several articles rely on a unique instrument based on theories that we could not place in a category. \cite{dower_etal2018} use religious polarization as an instrument for the frequency of unrest and argue that religious polarization could only impact collective action through its impact on representation in local institutions.

We wish to clarify that our reservations regarding instruments in this category are not primarily about theories themselves. As a design-based approach, the IV strategy requires \emph{specific} and \emph{precise} theories about the assignment process of the instruments and the exclusion restriction. We remain skeptical because many ``theory''-driven instruments, in our view, do not genuinely uphold these assumptions, often appearing to be developed in an ad hoc or post hoc manner.

The last category of instruments are based on econometric assumptions. This category includes what \cite{sovey2011instrumental} call ``Lags.'' These are econometric transformations of variables argued to constitute instruments. For example, \cite{lorentzen_etal2014} use a measure of the independent variable from eight years earlier to mitigate endogeneity concerns. Another example is shift-share ``Bartik" instruments based. For example, \cite{baccini2021}  use the interaction between job shares in specific industries and national employment changes to study the effect of manufacturing layoffs on voting.  The number of articles relying on econometric techniques, including flawed empirical tests (such as regressing $y$ on $d$ and $z$ and checking if the coefficient of $z$ is significant), has decreased.

\section{Replication Procedure and Results} \label{sec:repl}

In this section, we describe our replication procedure and report the main findings.

\paragraph*{Procedure.} For each paper, we select the main IV
specification that plays a central role in supporting a main claim in
the paper; it is either referred to as the baseline specification or
appears in one of the main tables or figures.  Focusing on this
specification, our replication procedure involves the following steps.
First, we compute the first-stage partial $F$-statistic based on (1)
classic analytic SEs, (2) Huber White heteroskedastic-robust SEs, (3) cluster-robust SEs (if applicable and based on the original specifications), and (4) bootstrapped SEs.%
\footnote{They are calculated by $F_{boot} =
\hat\tau_{2SLS}'{\hat{\mathbb{V}}_{boot}(\hat\tau_{2SLS})}^{-1}\hat\tau_{2SLS}/p_{z}$, where $p_{z}$
is the number of IVs and $\hat{\mathbb{V}}_{boot}(\hat\tau_{2SLS})$ is the
estimated variance-covariance matrix based on a nonparametric bootstrap
procedure, in which we repeatedly sample the rows of the data matrix
with replacement. If the data have a clustered structure, we use cluster-bootstrapping instead \parencite{Colin_Cameron2015-wp,Esarey2019-qt}. We include $F_{boot}$ as a reference to the classic $F$ and effective $F$. In Section A2 of the SM, we compare the five types of $F$-statistics and show that the effective $F$ and $F$ based on bootstrapping are usually more conservative (smaller) than other $F$-statistics.} %
We also calculate $F_{\texttt{Eff}}$. 

We then replicate the original IV result using the 2SLS estimator and apply four different inferential procedures. First, we make inferences based on analytic SEs, including robust SEs or cluster-robust SEs (if applicable). Additionally, we use two nonparametric bootstrap procedures, as described in Section 2, \emph{bootstrap-c} and \emph{bootstrap-t}. For specifications with only a single instrument, we also employ the $tF$ procedure proposed by \cite{Lee2020-mi}, using the 2SLS $t$-statistic and first-stage $F$-statistic based on analytic SEs accounting for the originally specified clustering structure. Finally, we conduct an AR procedure and record the $p$-values and CIs.

We record the point estimates, SEs (if applicable), 95\% CIs, and $p$-values for each procedure (the point estimates fully replicate the reported estimates in the original articles and are the same across all procedures). In addition, we estimate a naïve OLS model by regressing the outcome variable on the treatment and covariates, leaving out the instrument. We calculate the ratio between the magnitudes of the 2SLS and OLS estimates, as well as the ratio of their analytic SEs. We also record other useful information, such as the number of observations, the number of clusters, the types of instruments, the methods used to calculate SEs or CIs, and the rationale for each paper's IV strategy. Our replication yields the following three main findings.

\begin{figure}[!h]
\begin{minipage}[b]{.48\textwidth}
\begin{center}
   \includegraphics[width=\textwidth]{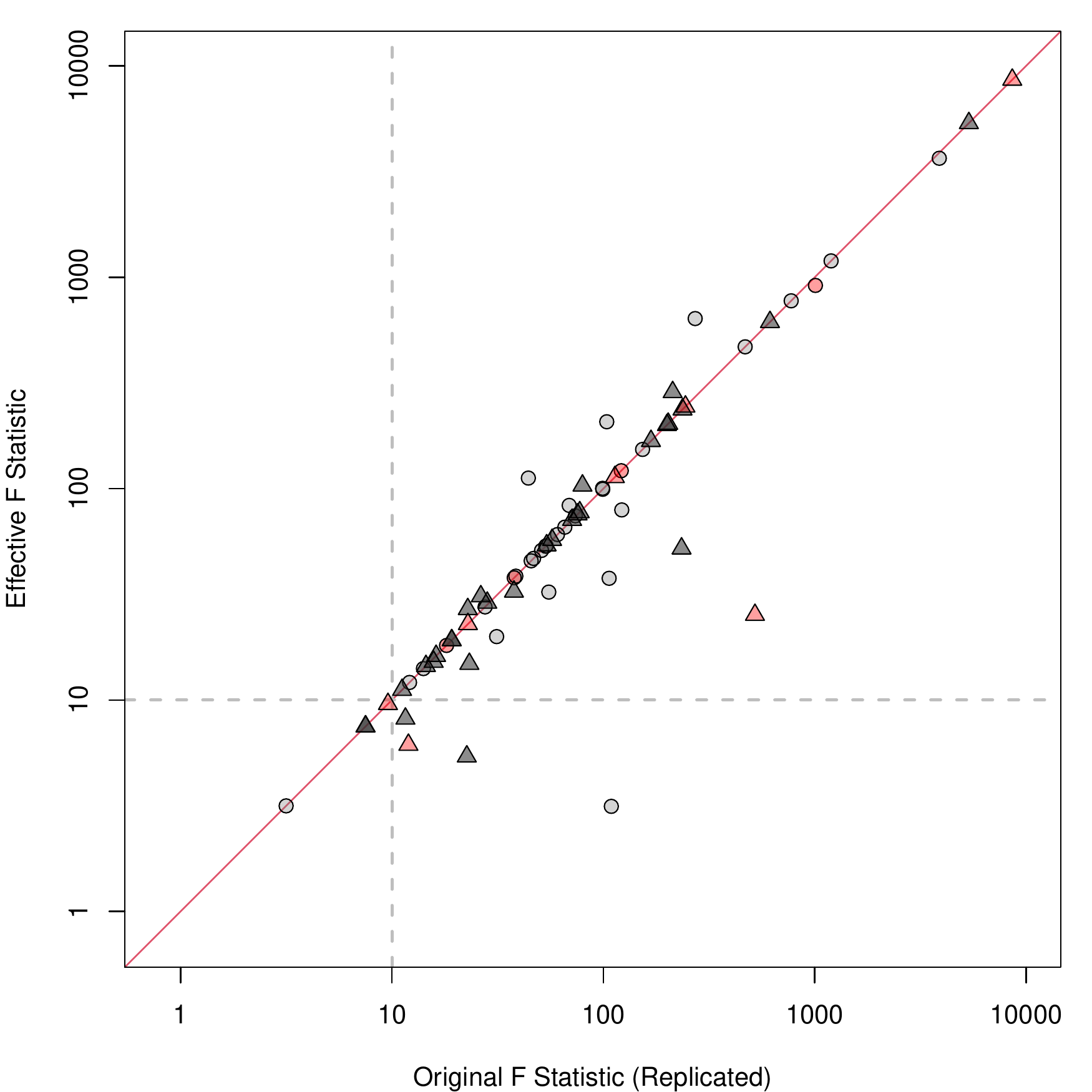}
    \subcaption{Original $F$ vs. Effective $F$}
    \label{fig:figure1}
    \end{center}
\end{minipage}
\hfill
\begin{minipage}[b]{.48\textwidth}
\begin{center}
    \includegraphics[width=\textwidth]{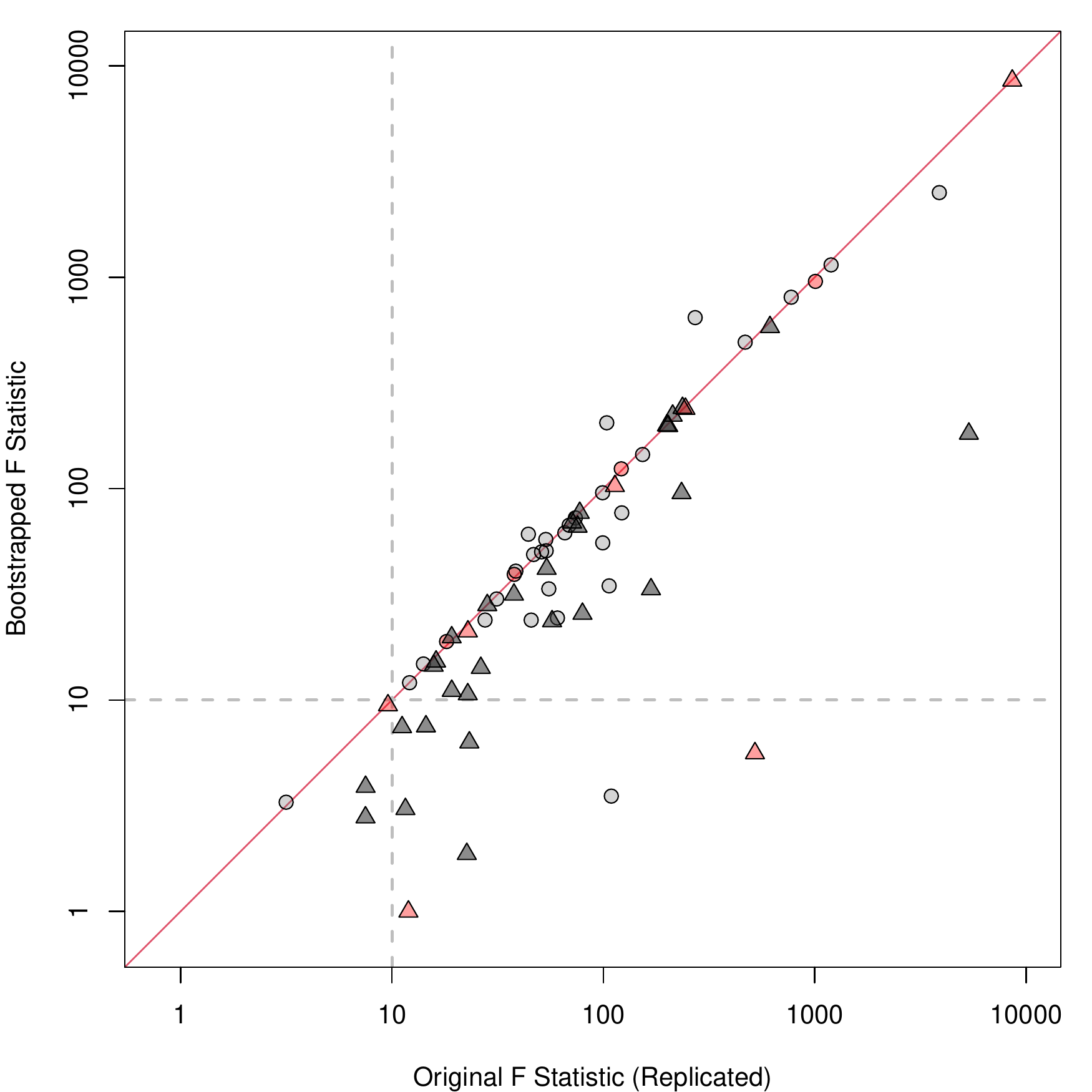}
    \subcaption{Original $F$ vs. Bootstrapped $F$}pip install arxiv-latex-cleaner
    \label{fig:figure2}
 \end{center}
 \end{minipage}
 \caption{Original vs. effective and bootstrapped $F$. Circles represent applications without a clustering structure and triangles represent applications with a clustering structure. Studies that do not report $F$-statistic are painted in red. The original $F$-statistics are obtained from the authors' original model specifications and choices of variance estimators in the 2SLS regressions. They may differ from those reported in the articles because of misreporting.}
  \label{fig:fstat}
\end{figure}

\paragraph*{Finding 1. The first-stage partial $F$-statistic.} Our first finding regards the strengths of the instruments. To our surprise, among the 70 IV designs, 12 (17\%) do not report this crucial statistic despite its key role in justifying the validity of an IV design. Among the remaining 58 studies that report $F$-statistic, 9 (16\%) use classic analytic SEs, thus not adjusting for potential heteroskedasticity or clustering structure. In Figure~\ref{fig:fstat}, we plot the replicated first-stage partial $F$-statistic based on the authors' original model specifications and choices of variance estimators on the x-axis against (a) effective $F$-statistic or (b) bootstrapped $F$-statistic on the y-axis, both on a logarithmic scale.\footnote{We use the replicated $F$-statistics instead of the reported ones because some authors either do not report or misreport their $F$-statistics (see SM for a comparison between the reported and replicated $F$-statistics).}

In the original studies, the authors used various SE estimators, such as classic SEs, robust SEs, or cluster-robust SEs. As a result, the effective $F$ may be larger or smaller than the original ones. However, a notable feature of Figure~\ref{fig:fstat} is that when a clustering structure exists, the original $F$-statistic tends to be larger than the effective $F$ or bootstrapped $F$. When using the effective $F$ as the benchmark, 8 studies (11\%) have $F_{\texttt{Eff}}<10$. This number increases to 12 (17\%) when the bootstrapped $F$-statistic is used. The median first-stage $F_{\texttt{Eff}}$ statistic is higher in experimental studies compared to non-experimental ones (67.7 versus 53.5). It is well known that failing to cluster the SEs at appropriate levels or using the analytic cluster-robust SE with too few clusters can lead to an overstatement of statistical significance \parencite{cameron2008bootstrap}. However, this problem has received less attention when evaluating IV strength using the $F$-statistic.%
\footnote{\cite{abadie2020sampling} and \cite{Abadie2022} delineate the differences between a traditional \emph{sampling-based} view, where clustering arises from a two-stage sampling process (sampling clusters, then units within them), and a \emph{design-based} view, where clustering stems from the clustered nature of treatment assignment. The key takeaway from both papers is the importance of clustering at the unit of randomization. They argue that finite-population standard errors, rooted in the design-based perspective, can be tighter than conventional cluster-robust errors. Given that the exact design is often unknown in many political science observational studies, clustering where the instrument is likely assigned offers a more reliable approach for valid inference. In the replication exercise, however, we cluster SEs according to the levels specified by the original authors.}

\begin{figure}[!ht]
\begin{minipage}[b]{0.48\textwidth}
    \centering
    \includegraphics[width=\textwidth]{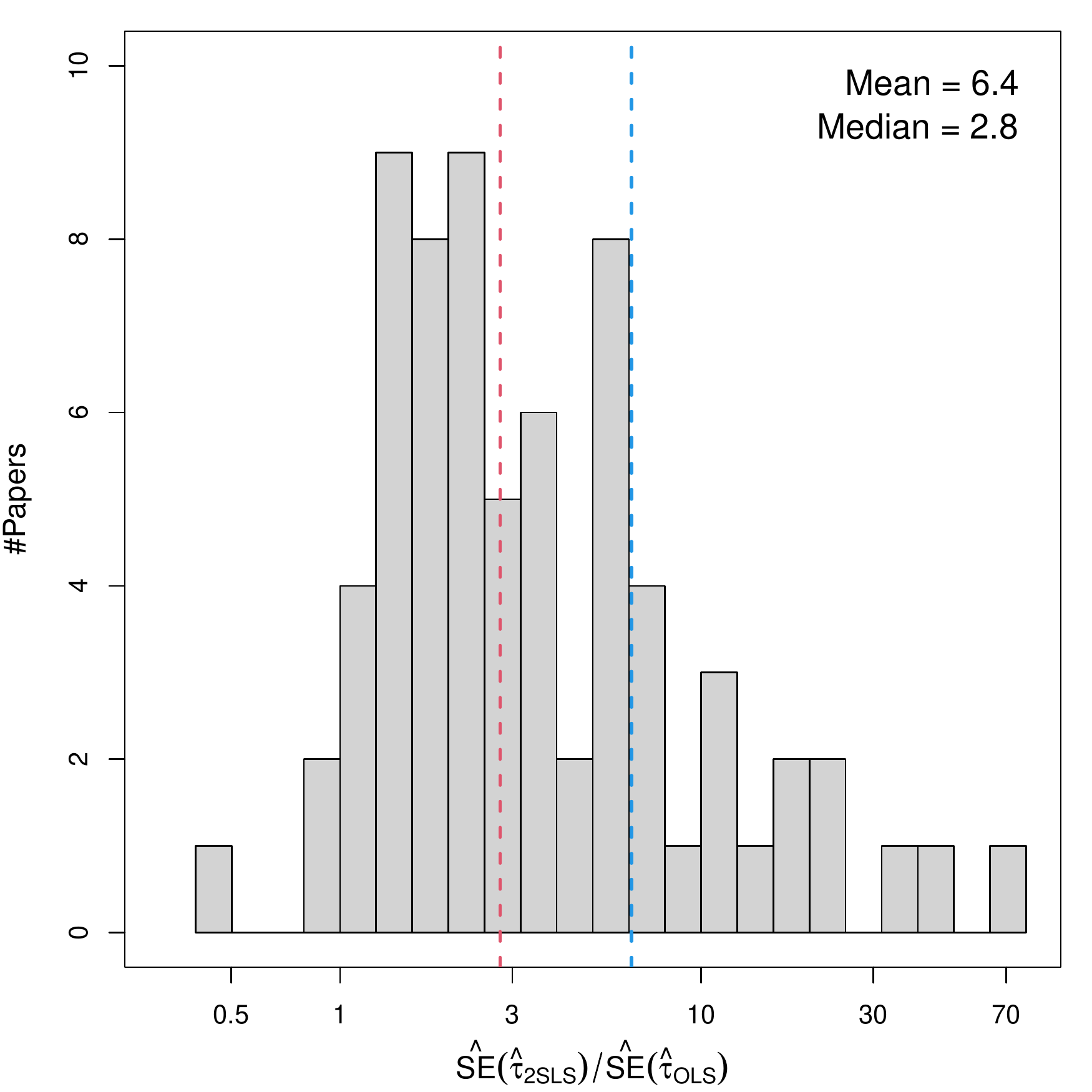}
    \subcaption{Distribution of Ratio of 2SLS and OLS SEs}
\end{minipage}
\hfill
\begin{minipage}[b]{0.48\textwidth}
    \centering
    \includegraphics[width=\textwidth]{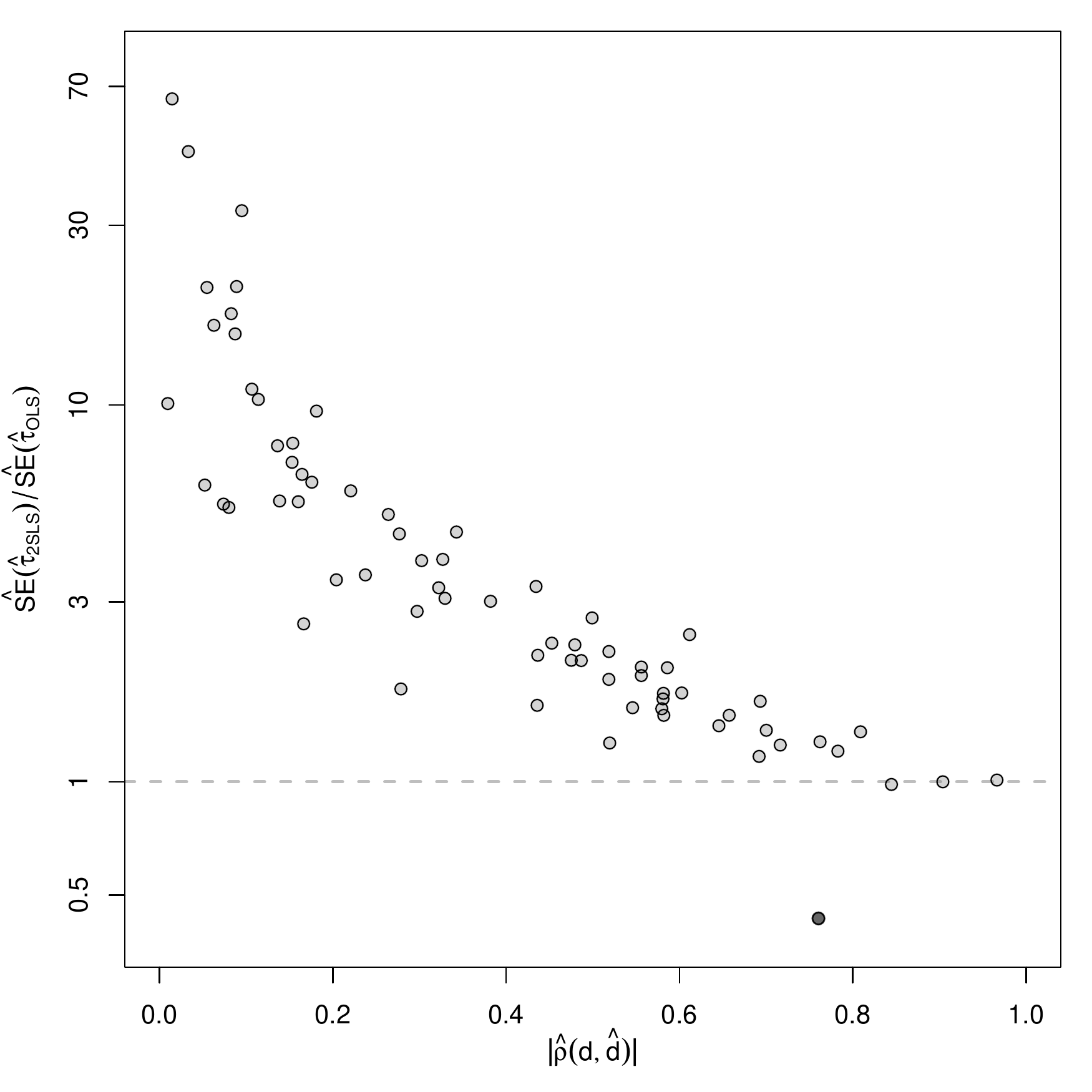}
    \subcaption{First Stage and the Ratio}
\end{minipage}\vspace{0.5em}
\caption{Comparison of 2SLS and OLS analytic SEs. Subfigure (a) shows the distribution of the ratio between $\hat{SE}(\hat\tau_{2SLS})$ and $\hat{SE}(\hat\tau_{OLS})$, both obtained analytically. Subfigure (b) plots the relationship between the absolute values of $\hat\rho(d, \hat{d})$, the estimated correlational coefficient between $d$ and $\hat{d}$, and the ratio (on a logarithmic scale). In one study, the analytic $\hat{SE}(\hat\tau_{2SLS})$ is much smaller than $\hat{SE}(\hat\tau_{OLS})$; we suspect that the former severely underestimates the true SE of the 2SLS estimate, likely due to a clustering structure.}\label{fig:se.ratio}
\end{figure}

\paragraph*{Finding 2. Inference.} Typically, 2SLS estimates have higher uncertainties than OLS estimates. Figure~\ref{fig:se.ratio} reveals that the 2SLS estimates in the replication sample are in general much less precise than their OLS counterparts, with the median ratio of the analytic SEs equal to 3.8. This ratio decreases as the strength of the instrument, measured by the estimated correlation coefficient between the treatment and predicted treatment $|\hat\rho(d, \hat{d})|$, increases. This is not surprising because $\hat\rho(d, \hat{d})^2 = R_{dz}^2$, the first-stage partial $R$-squared. However, one important implication of large differences in SEs is that to achieve comparable levels of statistical significance, 2SLS estimates often need to be at least three times larger than OLS estimates---not to mention that $t$-testing based on analytical SEs for 2SLS coefficients is often overly optimistic. This difference in precision sets the stage for potential publication bias and $p$-hacking.

\begin{figure}[!ht]
\begin{minipage}[b]{0.48\textwidth}
    \centering
    \includegraphics[width=\textwidth]{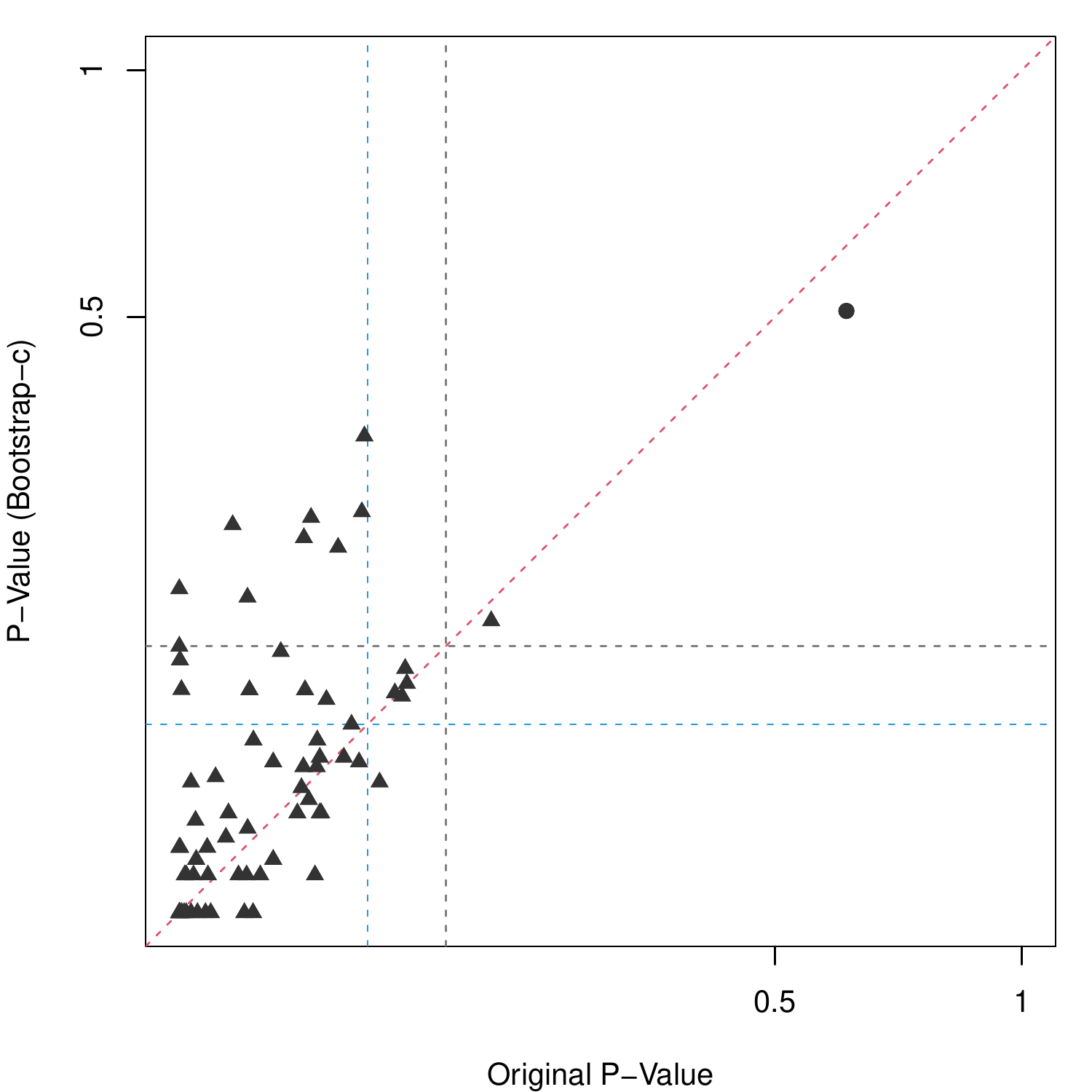}
    \subcaption{Bootstrap-c Method}
\end{minipage}
\hfill
\begin{minipage}[b]{0.48\textwidth}
    \centering
    \includegraphics[width=\textwidth]{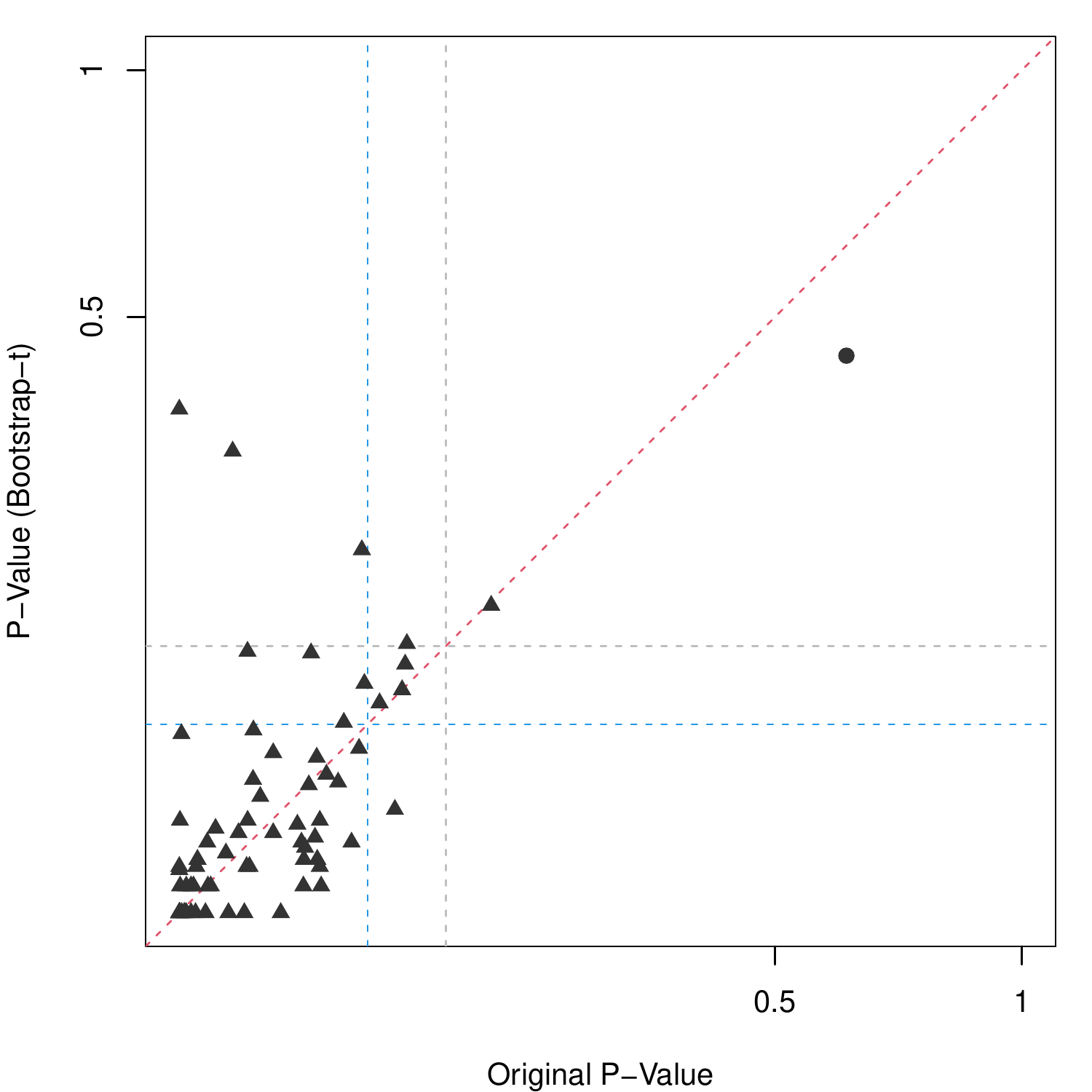}
    \subcaption{Bootstrap-t Method}
\end{minipage}\vspace{0.5em}
\begin{minipage}[b]{0.48\textwidth}
    \centering
    \includegraphics[width=\textwidth]{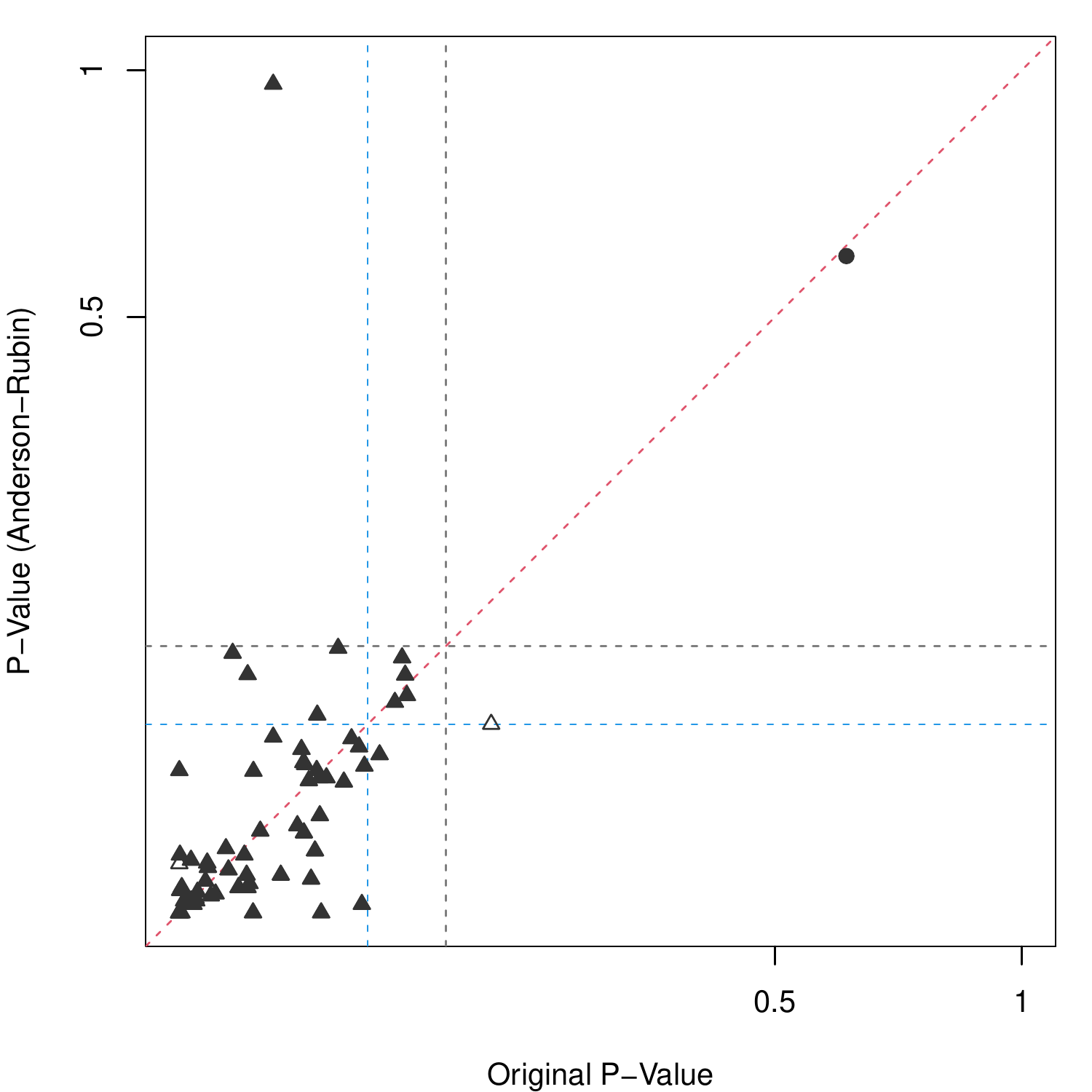}
    \subcaption{Anderson-Rubin}
\end{minipage}
\hfill
\begin{minipage}[b]{0.48\textwidth}
    \centering
    \includegraphics[width=\textwidth]{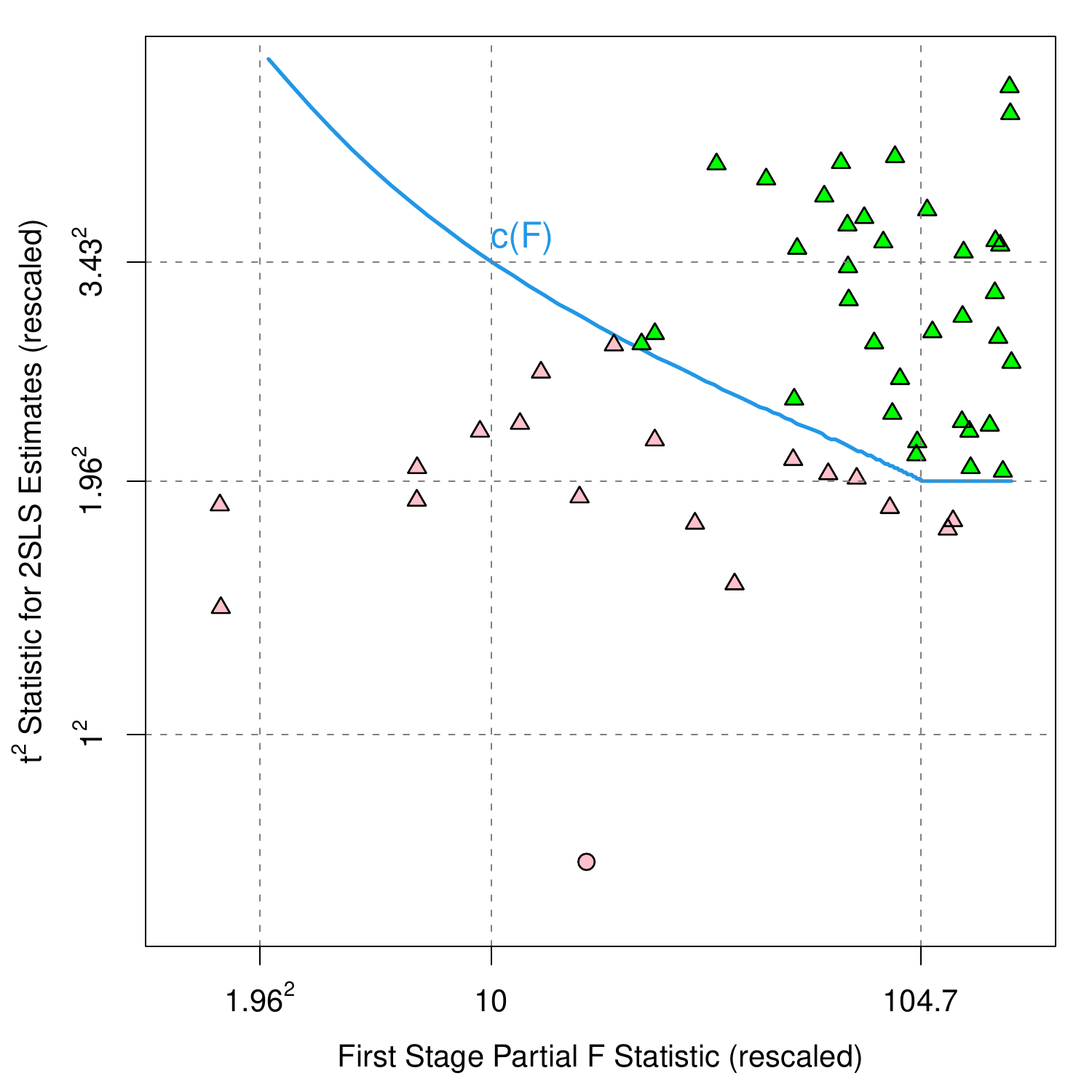}
    \subcaption{$tF$ Procedure}
\end{minipage}
\caption{Alternative inferential methods. In subfigures (a)-(c), we compare original $p$-values to those from alternative inferential methods, testing against the null that $\tau = 0$. Both axes use a square-root scale. Original $p$-values are adapted from original articles or calculated using standard-normal approximations of $z$-scores. Solid circles represent \cite{arias2019large}, where authors argue for a null effect using IV strategy. \emph{Bootstrap-c} and \emph{-t} represent percentile methods based on 2SLS estimates and $t$-statistics, respectively, using original model specifications. Hollow triangles in subfigure (c) indicate unbounded 95\% CIs from the AR test using the inversion method. Subfigure (d) presents $tF$ procedure results from 54 single instrument designs. Green and red dots represent studies remaining statistically significant at the 5\% level using the $tF$ procedure and those that don't, respectively. Subfigures (a)-(c) are inspired by Figure 3 in \cite{Young2022}, and subfigure~(d) by Figure~3 in \cite{Lee2020-mi}.}\label{fig:inference}
\end{figure}

Next, we compare the reported and replicated $p$-values for the null hypothesis of no effect. For studies that do not report a $p$-value, we calculate it based on a standard normal distribution using the reported point estimates and SEs. The replicated $p$-values are based on (1) \emph{bootstrap-c}, (2) \emph{bootstrap-t}, and (3) the AR procedure. Since we can exactly replicate the point estimates for the articles in the replication sample, the differences in $p$-values are the result of the inferential methods used. Figure~\ref{fig:inference}(a)-(c) plot reported and replicated $p$-values, from which we observed two patterns. First, most of the reported $p$-values are smaller than 0.05 or 0.10, the conventional thresholds for statistical significance. Second, consistent with \cite{Young2022}'s finding, our replicated $p$-values based on the bootstrap methods or AR procedure are usually bigger than the reported $p$-value (exceptions are mostly caused by rounding errors), which are primarily based on $t$ statistics calculated using analytic SEs. Using the AR test, we cannot reject the null hypothesis of no effect at the 5\% level in 12 studies (17\%), compared with 7 (10\%) in the original studies. The number increases to 13 (19\%) and 19 (27\%) when we use $p$-values from the \emph{bootstrap-t} and \emph{-c} methods. Note that very few articles we review utilize inferential procedures specifically designed for weak instruments, such as the AR test (2 articles), the conditional likelihood-ratio test \parencite{Moreira2003-oj} (1 paper), and confident sets \parencite{Mikusheva2006-lk} (none).

We also apply the $tF$ procedure to 54 studies that use single IVs using $F_{\texttt{Eff}}$ statistics and $t$ statistics based on robust or cluster-robust SEs. Figures~\ref{fig:inference}(d) shows that 19 studies (35\%) are not statistically significant at the 5\% level, and 7 studies (13\%) deemed statistically significant when using the conventional fixed critical values for the $t$-test become statistically insignificant using the $tF$ procedure, indicating that overly optimistic critical values due to weak instruments also contribute to overestimation of statistical power, but not as the primary factor. These results suggest that both weak instruments and non-i.i.d. errors have contributed to overstatements of power in IV studies in political science.

\FloatBarrier

\begin{figure}[!ht]
\begin{minipage}[b]{.48\textwidth}
   \begin{center}
   \includegraphics[width=\textwidth]{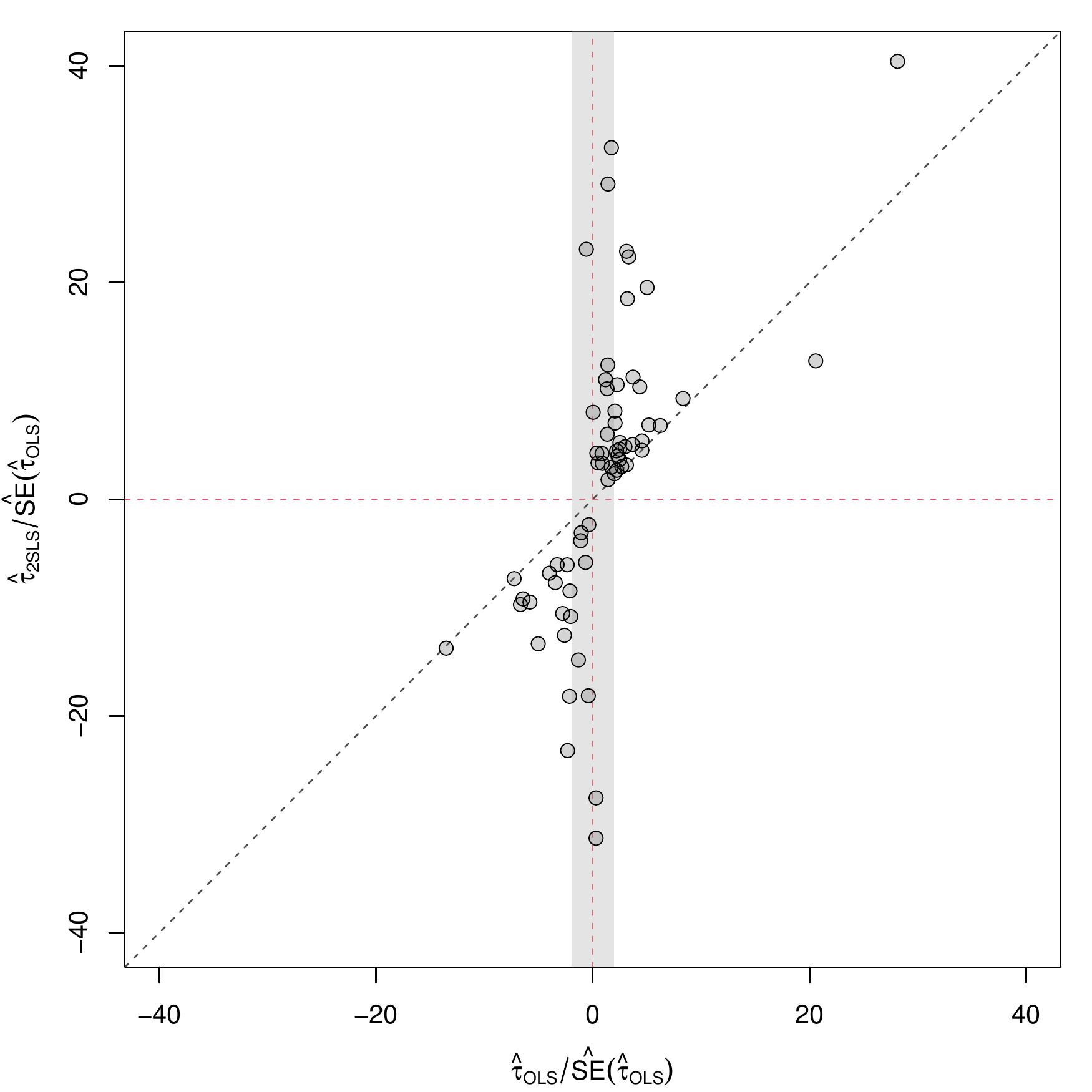}
   \subcaption{OLS vs 2SLS Coefficients}
   \end{center}
 \end{minipage}\hfill
 \begin{minipage}[b]{.48\textwidth}
   \begin{center}
   \includegraphics[width=\textwidth]{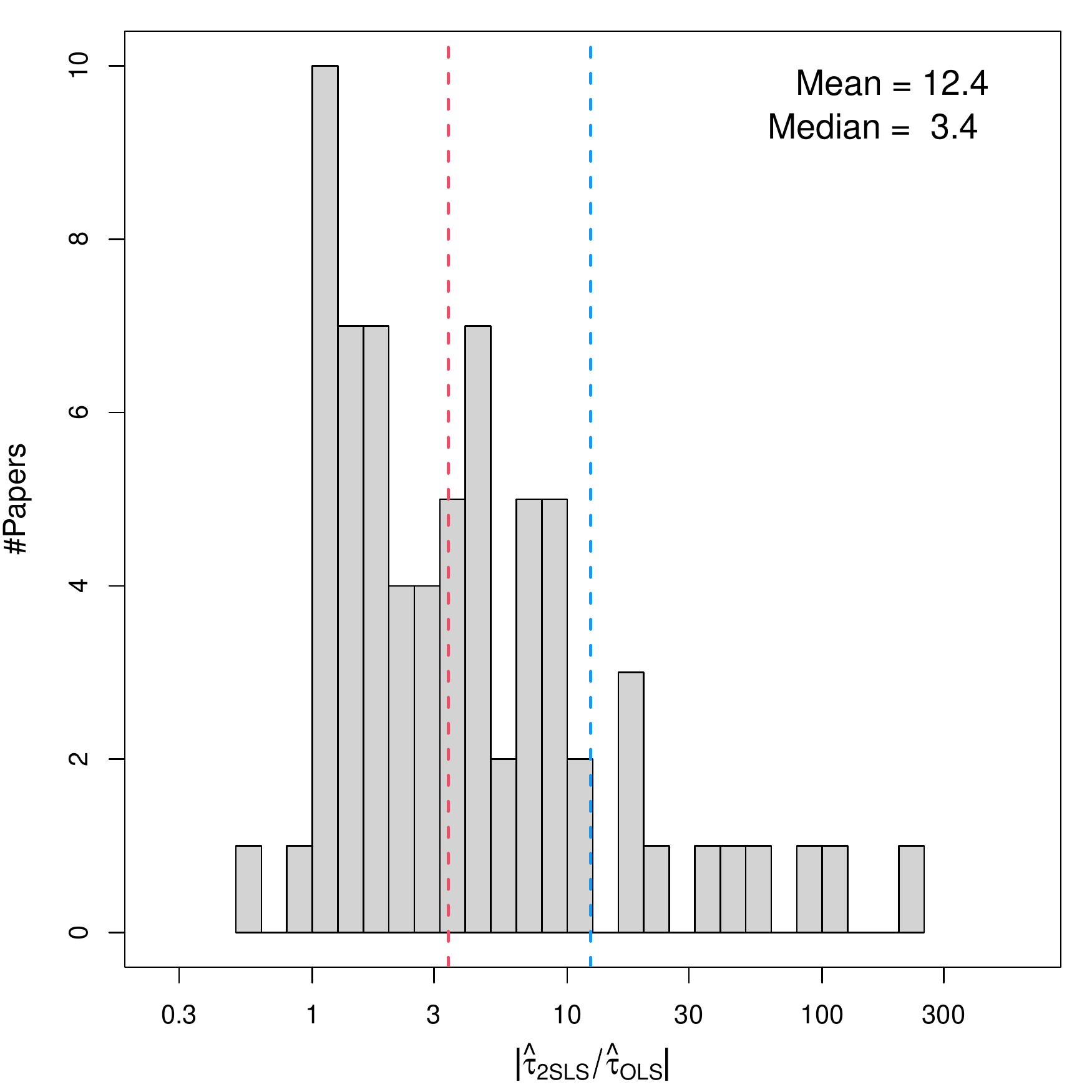}
   \subcaption{Distribution of Ratio of 2SLS and OLS Coefficients}
   \end{center}
\end{minipage}
\begin{minipage}[b]{.48\textwidth}
   \begin{center}
   \includegraphics[width=\textwidth]{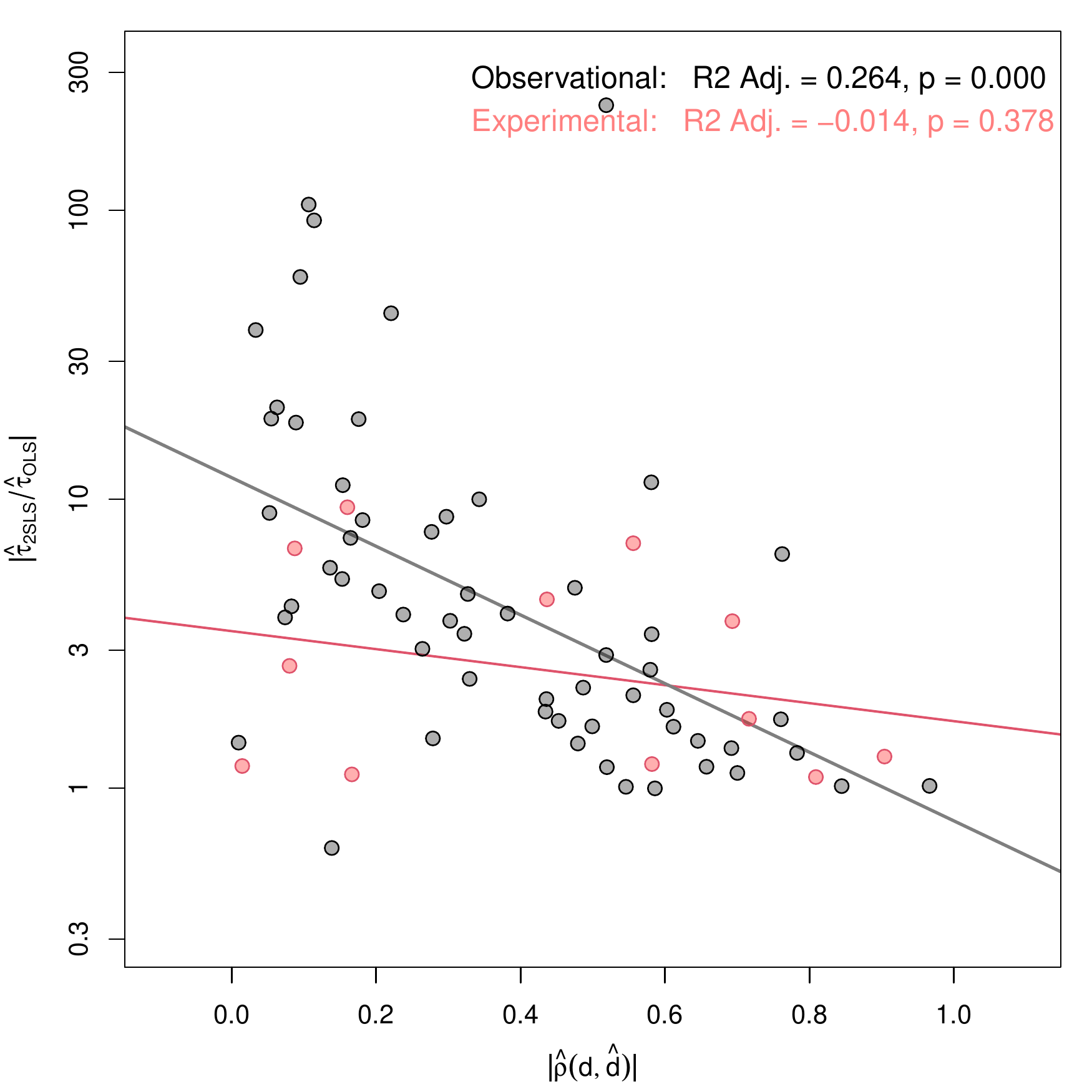}
   \subcaption{First Stage and the Ratio: Full Sample}
   \end{center}
 \end{minipage}\hfill
 \begin{minipage}[b]{.48\textwidth}
   \begin{center}
   \includegraphics[width=\textwidth]{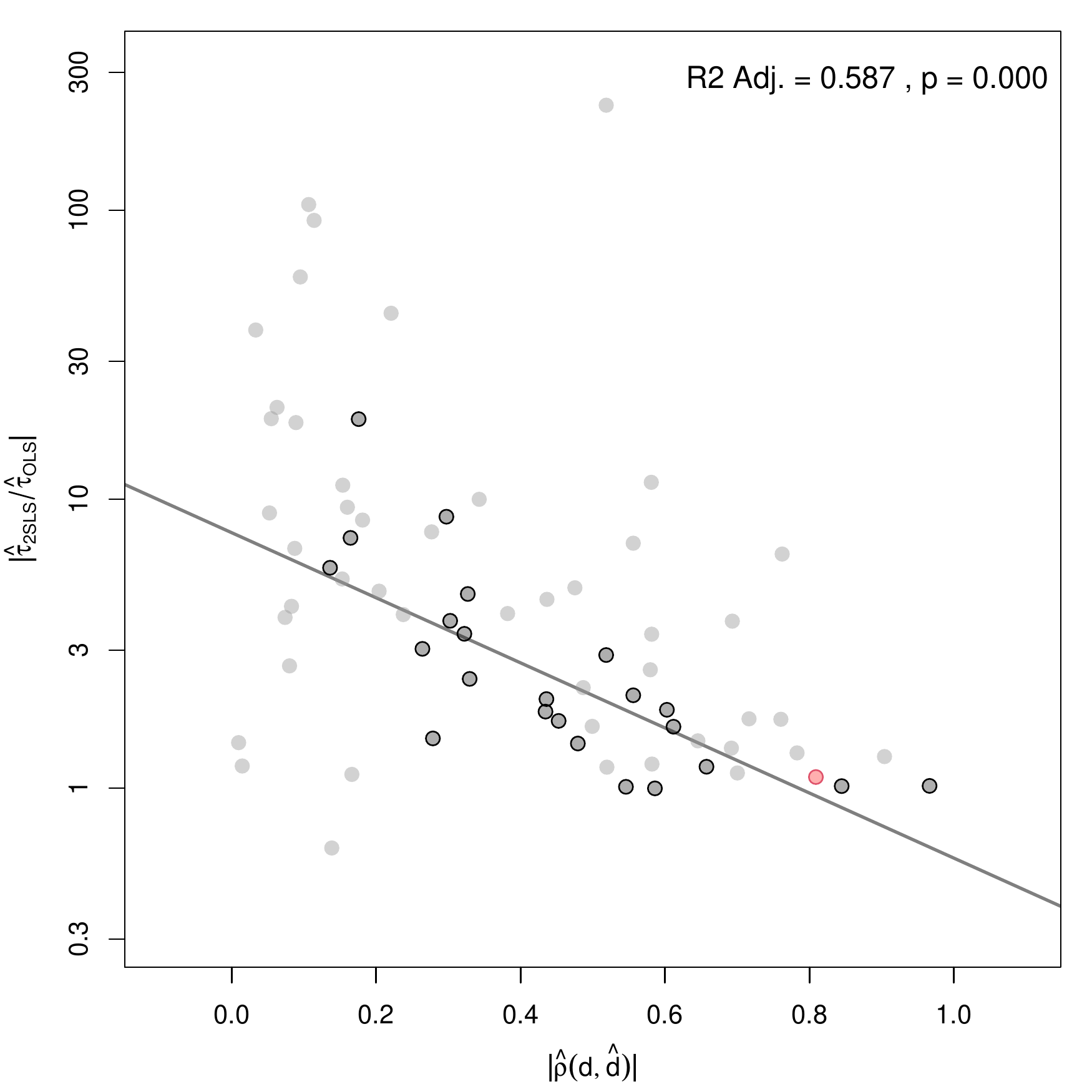}
   \subcaption{First Stage and Ratio: Subsample}
   \end{center}
\end{minipage}
\caption{Relationship between OLS and 2SLS estimates. In subfigure (a), both axes are normalized by reported OLS SE estimates with the gray band representing the $[-1.96, 1.96]$ interval. Subfigure (b) displays a histogram of the logarithmic magnitudes of the ratio between reported 2SLS and OLS coefficients. Subfigures (c) and (d) plot the relationship between $|\hat\rho(d,\hat{d})|$ and the ratio of 2SLS and OLS estimates. Gray and red circles represent observational and experimental studies, respectively. Subfigure (d) highlights studies with statistically significant OLS results at the 5\% level, claimed as part of the main findings.}\label{fig:ratio}
\end{figure}

\paragraph*{Finding 3. 2SLS-OLS discrepancy.} Finally, we investigate the relationship between the 2SLS estimates and naïve OLS estimates. In Figure~\ref{fig:ratio}(a), we plot the 2SLS coefficients against the OLS coefficients, both normalized using reported OLS SEs. The shaded area indicates the range beyond which the OLS estimates are statistically significant at the 5\% level. It shows that for most studies in our sample, the 2SLS estimates and OLS estimates share the same direction and that the magnitudes of the former are often much larger than those of the latter. Figure~\ref{fig:ratio}(b) plots the distribution of the ratio between the 2SLS and OLS estimates (in absolute terms). The mean and median of the absolute ratios are 12.4 and 3.4, respectively. In fact, in all but two designs (97\%), the 2SLS estimates are bigger than the OLS estimates, consistent with \cite{jiang2017have}'s finding based on finance research. While it is theoretically possible for most OLS estimates in our sample to be biased towards zero, only 21\% of the studies have researchers expressing their belief in downward biases of the OLS estimates. Meanwhile, 40\% of the studies consider the OLS results to be their main findings. The fact that researchers use IV designs as robustness checks for OLS estimates due to concerns of upward biases is apparently at odds with the significantly larger magnitudes of the 2SLS estimates.

\FloatBarrier

In Figure~\ref{fig:ratio}(c), we further explore whether the 2SLS-OLS discrepancy is related to IV strength, measured by $|\hat\rho(d, \hat{d})|$. We find a strong negative correlation between 
$|\hat{\tau}_{2SLS} / \hat{\tau}_{OLS}|$ and $|\hat\rho(d,\hat{d})|$ among studies using non-experimental instruments (grey dots). The adjusted $R^2$ is $0.264$, with $p = 0.000$. However, the relationship is much weaker among studies using experiment-generated instruments (red dots). The adjusted $R^2$ is $-0.014$ with $p = 0.378$. At first glance, this result may seem mechanical: as the correlation between $d$ and $\hat{d}$ increases, the 2SLS estimates naturally converge to the OLS estimates. However, the properties of the 2SLS estimator under the identifying assumptions do not predict the negative relationship (we confirm it in simulations in the SM), and such a relationship is not found in experimental studies. In Figure~\ref{fig:ratio}(d), we limit our focus to the subsample in which the OLS estimates are statistically significant at the 5\% level and researchers accept them as (part of) the main findings, and the strong negative correlation remains. 

Several factors may be contributing to this observed pattern, including (1) failure of the exogeneity assumption, (2) publication bias, (3) heterogeneous treatment effects, and (4) measurement error in $d$. As noted earlier, biases originating from endogenous IVs or exclusion restriction failures can be magnified by weak instruments, i.e., $\frac{|\text{Bias}_{IV}|}{|\text{Bias}_{OLS}|} = \left|\frac{\Covar{z, \varepsilon}\Var{d}}{ \Covar{d, \varepsilon}\Covar{z, d}}\right| = \frac{|\rho(z, \varepsilon)|}{|\rho(d, \varepsilon)|\cdot |\rho(d, \hat{d})|} \gg~1$. In addressing large IV-OLS estimate ratios, \cite{hahn2005estimation} suggest two explanations: it could stem from a bias in OLS or from a bias in IV due to violations of the exogeneity assumption.  Our empirical results, with particularly dubious IV to OLS estimate ratios in non-experimental studies, seem to align with the latter explanation.

Publication bias may have also played a significant role.  As shown in Figure~\ref{fig:se.ratio}, the variance of IV estimates increase as $|\hat\rho(d,\hat{d})|$ diminishes. If researchers selectively report only statistically significant results, or if journals have a tendency to publish such findings, it is not surprising that the discrepancies between IV and OLS estimates widen as the strength of the first stage declines, as shown in Figure~\ref{fig:ratio}(c)-(d). This is because 2SLS estimates often need to be substantially larger than OLS estimates to achieve statistical significance. This phenomenon is known as Type-M bias and has been discussed in psychology and sociology literature \parencite{gelman2014beyond, FeltonStewart2022}. Invalid instruments exacerbate this issue by providing ample opportunities for generating such large estimates.

Moreover, 30\% of the replicated studies in our sample mention heterogeneous treatment effects as a possible explanation for this discrepancy. OLS and 2SLS place different weights on covariate strata in the sample, and therefore if compliers, those whose treatment status is affected by the instrument, are more responsive to the treatment than the rest of the units in the sample, we might see diverging OLS and 2SLS estimates. Under the assumption that the exclusion restriction holds, this gap can be decomposed into covariate weight difference, treatment-level weight difference, and endogeneity bias components using the procedure developed in \cite{Ishimaru2021-ik}. In the SM, we investigate this possibility and find that it is highly unlikely that heterogeneous treatment effects \emph{alone} can explain the difference in magnitudes between 2SLS and OLS estimates we observe in the replication data, i.e.,  the variance in treatment effects needed for this gap is implausibly large.

Finally, IV designs can correct for downward biases due to measurement errors in $d$, resulting in $|\hat{\tau}_{2SLS} / \hat{\tau}_{OLS}| > 1$. If the measurement error is large, this can weaken the relationship between $d$ and $\hat{d}$, producing a negative correlation. We find it an unlikely explanation because only four articles (6\%) attribute their use of IV to measurement errors, and the negative correlation is even stronger when we focus solely on studies where OLS estimates are statistically significant and regarded as the main findings.

\begin{table}[!ht]
  \centering\small
  \caption{Summary of replication results}
  \label{tb:des}%
    \begin{tabular}{p{15em}ccc}\hline\hline
     & Experimental (12) & Observational (58) &  All (70) \\ \hline
    \multicolumn{3}{l}{\emph{Panel A: First-Stage $F$-statistic}}  & \\
    $\quad$Unreported &  .333   & .138  &  .171\\
    $\quad$Effective $F < 10$ &  .083 & .121 &  .114 \\
    $\quad$Bootstrapped $F < 10$ &  .167 & .172  &  .171 \\ 
    $\quad$Median effective $F$ &  67.7 & 53.5  &  53.5 \\ \\
    \multicolumn{3}{l}{\emph{Panel B: Inference for 2SLS Estimates}}  & \\
    $\quad$Original $p > 0.05$ &  .250  & .069 & .100 \\
    $\quad$AR $p > 0.05$ &  .417   & .121 & .171 \\
    $\quad$Bootstrap-c $p > 0.05$  &  .417   & .241 & .271 \\
    $\quad$Bootstrap-t $p > 0.05$ &   .333  & .155  & .186 \\
    $\quad$$tF$ procedure $p > 0.05$  &  .400  & .341 & .352  \\ \\
    \multicolumn{3}{l}{\emph{Panel C: 2SLS-OLS Relationship}} & \\
    $\quad$ $sign(\hat\tau_{2SLS}) = sign(\hat\tau_{OLS})$  &  1.00 & .914 & .929 \\
    $\quad\ |\hat\tau_{2SLS}/\hat\tau_{OLS}|> 1$ &  1.00 & .966  & .971 \\
    $\quad\ |\hat\tau_{2SLS}/\hat\tau_{OLS}|> 5$  &  .250  & .362  & .343 \\
    $\quad\ |\hat\tau_{2SLS}/\hat\tau_{OLS}|> 10$ &  .000   & .207 & .171 \\
    $\quad$Median $|\hat\tau_{2SLS}/\hat\tau_{OLS}|$ & 2.19 & 3.61  &  3.40 \\
    \hline
    \end{tabular}%
\end{table}

In Table~\ref{tb:des}, we present the main findings from our replication exercise. Observational studies, compared to experimental counterparts, generally have weaker first stages, often display larger increases in $p$-values when more robust inferential methods are used, and demonstrate bigger discrepancies between the 2SLS and OLS estimates. Based on these findings, we contend that a significant proportion of IV results based on observational data in political science either lack credibility or yield estimates that are too imprecise to offer insights beyond those provided by OLS regressions.

\section{Recommendations} \label{conc}

IV designs in experimental and observational studies differ fundamentally. In randomized experiments, the instruments' unconfoundedness is ensured by design, and researchers can address possible exclusion restriction violations at the design stage, e.g., by testing potential design effects through randomization \parencite[pp. 140-141]{Gerber2012-fr}. Practices like power analysis, placebo tests, and preregistration also help reduce the improper use of IVs. In contrast, observational IV designs based on ``natural experiments'' require detailed knowledge of the assignment mechanism, making them more complex and prone to issues \parencite{sekhon2012natural}.

Our findings suggest that using an IV strategy in observational settings is extremely challenging due to several reasons. First, truly random and strong instruments are rare and difficult to find. This is mainly because neither unconfoundedness nor the exclusion restriction is guaranteed by design, placing a greater burden of proof on researchers for the exogeneity assumption. Moreover, conducting placebo tests like the ZFS test for the exclusion restriction after data collection is not always feasible. Finally, increasing the sample size to achieve sufficient statistical power is often impractical. To prevent misuse of IVs in observational studies, we provide a checklist for researchers to consider when applying or contemplating an IV strategy with one endogenous treatment variable:

\begin{figure}[!ht]
    \vspace{-0.5em}
        \begin{minipage}{1\linewidth}{
        \begin{center}
        \includegraphics[width=0.8\textwidth]{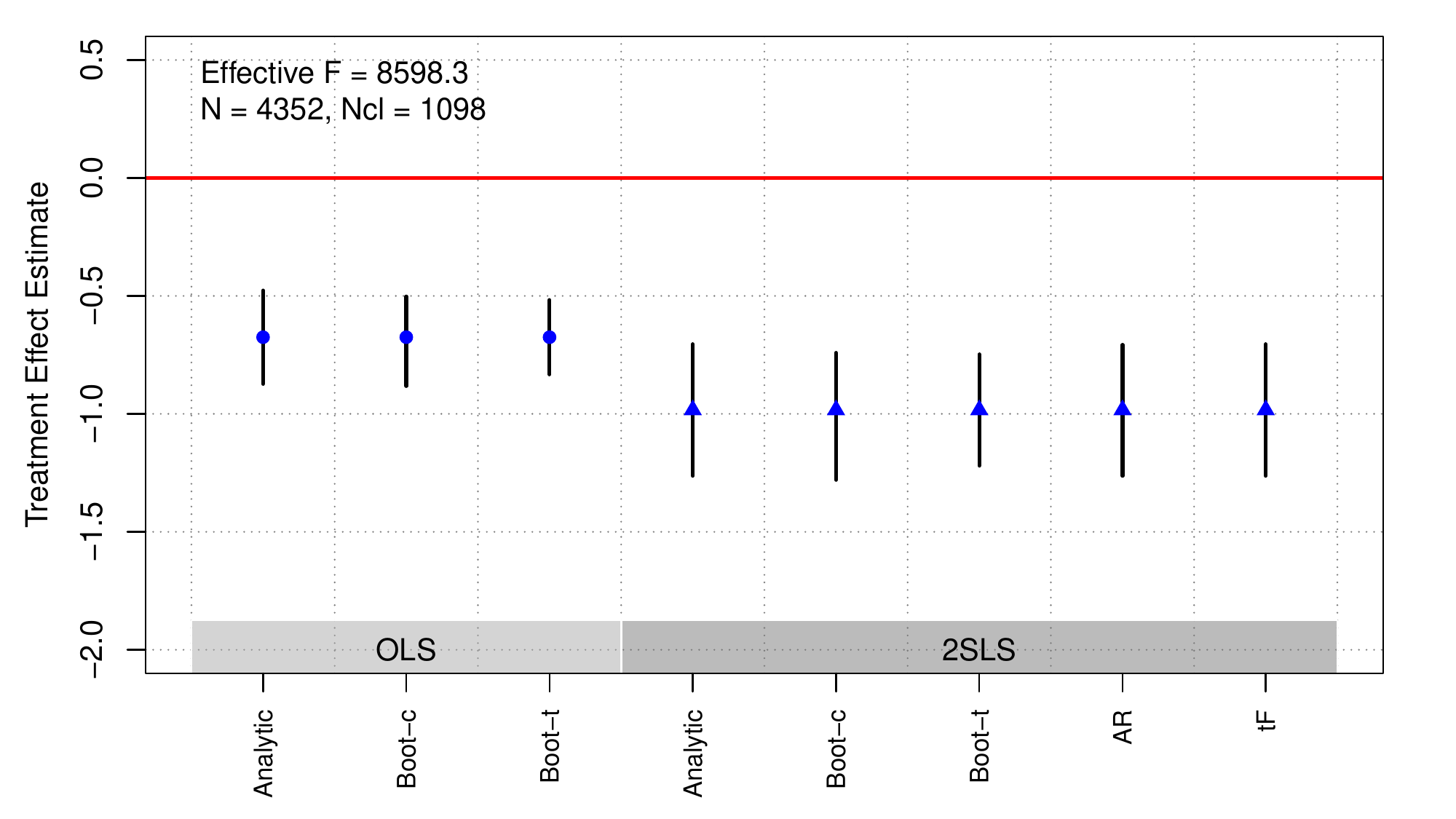}\hspace{0em}
        \end{center}\vspace{-1.5em}
        \caption{Replicated OLS and 2SLS estimates with 95\% CIs \parencite[][Table 5 column 1]{rueda2017}. The outcome is citizens' reports of voting buying. The treatment is the actual polling place size. The instrument is the size of the polling station predicted by the rules limiting the voters per polling station. The magnitude of the 2SLS estimate is slightly larger than that of the OLS estimate. Similar figures for each of the 70 IV designs are shown in the SM. This plot is made by \href{https://github.com/apoorvalal/ivDiag}{ivDiag}.}\label{fig:estimates}}
        \end{minipage}\vspace{-0.5em}
\end{figure}

\FloatBarrier

\begin{itemize}
    \item[] \textbf{Design}
    \item Prior to using an IV strategy, consider how selection bias may affect treatment effect estimates obtained through OLS. If the main concern is underestimating an already statistically significant treatment effect, an IV strategy may be unnecessary. 
    \item During the research design phase, consider whether the chosen instrument can realistically create random or quasi-random variations in treatment assignment while remaining excluded from the outcome equation.
    \item[] \textbf{Characterizing the first-stage}
    \item Calculate and report $F_{\texttt{Eff}}$ for the first stage, taking into account heteroscedasticity and clustering structure as needed. However, do not discard a design simply because $F_{\texttt{Eff}}< 10$.
    \item If both $d$ and $z$ are continuous, we recommend plotting $d$ against its predicted values, $\hat{d}$, after accounting for covariates. Alternatively, plot both $d$ and $\hat{d}$ against specific covariates that serve as the foundation for the rules used to derive the instruments. These visualizations are useful for detecting outliers and gaining insights into the sources of exogenous variation.
    \item[] \textbf{Hypothesis testing and inference}
    \item \textit{Option 1. $t$-test with $F_{\texttt{Eff}}$ pretesting.} If $F_{\texttt{Eff}} < 10$, choose Options 2 or 3. Utilize conservative methods like \emph{bootstrap-t} and \emph{bootstrap-c} if outliers or group structures are present.
    \item \textit{Option 2. $tF$ procedure.} For single treatment and instrument cases, adjust $t$-test critical values based on $F_{\texttt{Eff}}$.
    \item \textit{Option 3. Direct testing.} Apply weak-instrument-robust procedures, such as the AR test.
    \item[] \textbf{Communicating your findings}
    \item Present OLS and IV estimates alongside CIs from various inferential methods in a graphical format, like in Figure~\ref{fig:estimates}. These CIs may not concur on statistical significance, but they collectively convey the findings' robustness to different inferential approaches. In addition, they present the degree of uncertainty in both OLS and IV estimates in an intuitive manner.
    \item Remember to report first-stage and reduced-form estimation results, including 95\% CIs for coefficients, as they offer insight into both instrument strength and statistical power.
    \item[] \textbf{Additional diagnostics}
    \item If you expect the OLS results to be upward biased, be concerned if the 2SLS estimator yields much larger estimates.
    \item If there is good reason to believe that treatment effects on compliers are significantly larger in magnitude than those on non-compliers, explain this through profiling of these principal strata \parencite{Abadie2003-sl, Marbach2020-ts}.
    \item If it is possible to identify ``never takers'' or a subset of them, conduct a placebo test by estimating the effect of the instruments on the outcome in this ZFS sample. Using results from the ZFS test, obtain local-to-zero IV estimates and CIs and compare them to the original estimates. Section A3 of the SM provides detailed explanations and an empirical example.
    \item Conduct a sensitivity analysis as proposed by \cite{cinelli2022omitted}.
\end{itemize}\medskip

We provide an \texttt{R} package, \href{https://github.com/apoorvalal/ivDiag}{ivDiag}, to implement our recommended procedures. Stata tutorials for carrying out these procedures are also available on the corresponding author's website. Our aim is to address concerns regarding IVs in social science research and improve the quality of estimation and inference, especially for non-experimental IV designs.

\section*{Acknowledgements}
We thank Te Bao, Thomas Cao, Albert Chiu, Daniel Chen, Gary Cox, Charles Crabtree, Ted Enamorado, Hanming Fang, Avi Feller, Don Green, Justin Grimmer, Anna Grzymala-Busse, Jens Hainmueller, Guido Imbens, David Laitin, Adeline Lo, Justin McCrary, Jacob Montgomery, Doug Rivers, Henrik Sigstad, Brandon Stewart, Arthur Yu, and Xiang Zhou, as well as seminar participants at Stanford University, Washington University in St. Louis, APSA 2021, and Polmeth 2021, for their extremely helpful comments. We are also deeply grateful to four anonymous reviewers for PA and Editor Jeff Gill, as well as two anonymous reviewers for APSR---their invaluable suggestions have improved this paper significantly.

\section*{Data Availability Agreement}
Replication data and code for this article have been published at Harvard Dataverse at \url{https://doi.org/10.7910/DVN/MM5THZ}.

\section*{Supplementary Material}
For supplementary material accompanying this paper, please visit \url{https://doi.org/10.xxxx/ pan.20xx.xx}.

\singlespacing
\renewcommand{\mkbibnamefamily}[1]{\textsc{#1}}
\printbibliography

@article{chiu2023what,
  title={What To Do (and Not to Do) with Causal Panel Analysis under Parallel Trends: Lessons from a Large Reanalysis Study},
  author={Chiu, Albert and Lan, Xingchen and Liu, Ziyi and Xu, Yiqing},
  year={2023},
  url={https://ssrn.com/abstract=4490035}
}

@article{abadie2020sampling,
  title={Sampling-based versus Design-based Uncertainty in Regression Analysis},
  author={Abadie, Alberto and Athey, Susan and Imbens, Guido W and Wooldridge, Jeffrey M},
  journal={Econometrica},
  volume={88},
  number={1},
  pages={265--296},
  year={2020}
}

@article{Abadie2022,
  title={When Should You Adjust Standard Errors for Clustering?},
  author={Abadie, Alberto and Athey, Susan and Imbens, Guido W and Wooldridge, Jeffrey M},
  journal={The Quarterly Journal of Economics},
  volume={138},
  number={1},
  pages={1--35},
  year={2023}
}

@article{angrist2023one,
  title={One Instrument to Rule Them All: The Bias and Coverage of Just-ID IV},
  author={Angrist, Joshua and Koles{\'a}r, Michal},
  journal={Journal of Econometrics},
  year={2023},
  pages = {105398}
}

@article{hahn2005estimation,
  title={Estimation with Valid and Invalid Instruments},
  author={Hahn, Jinyong and Hausman, Jerry},
  journal={Annales d'Economie et de Statistique},
  pages={25--57},
  year={2005},
  publisher={JSTOR}
}

@article{stommes2023reliability,
  title={On the Reliability of Published Findings Using the Regression Discontinuity Design in Political Science},
  author={Stommes, Drew and Aronow, PM and S{\"a}vje, Fredrik},
  journal={Research \& Politics},
  volume={10},
  number={2},
  pages={1-12},
  year={2023},
  publisher={SAGE Publications Sage UK: London, England}
}

@article{Abadie2003-sl,
	title        = {Semiparametric Instrumental Variable Estimation of Treatment Response Models},
	journal      = {Journal of Econometrics},
	volume       = 113,
	number       = 2,
	pages        = {231--263},
	year         = 2003,
	author       = {Alberto Abadie}
}

@article{anderson1949estimation,
	title        = {Estimation of the Parameters of a Single Equation in a Complete System of Stochastic Equations},
	author       = {Anderson, Theodore W and Rubin, Herman},
	journal      = {The Annals of Mathematical Statistics},
	volume       = 20,
	number       = 1,
	pages        = {46--63},
	year         = 1949
}

@article{anderson1982evaluation,
	author       = {T. W. Anderson and Naoto Kunitomo and Takamitsu Sawa},
	journal      = {Econometrica},
	number       = 4,
	pages        = {1009--1027},
	title        = {Evaluation of the Distribution Function of the Limited Information Maximum Likelihood Estimator},
	volume       = 50,
	year         = 1982
}

@article{andrews2009validity,
	author       = {Donald W. K. Andrews and Patrik Guggenberger},
	journal      = {Econometric Theory},
	number       = 3,
	pages        = {669--709},
	publisher    = {Cambridge University Press},
	title        = {Validity of Subsampling and ``Plug-in Asymptotic" Inference for Parameters Defined by Moment Inequalities},
	volume       = 25,
	year         = 2009
}

@article{andrews2019weak,
	author       = {Andrews, Isaiah and Stock, James H. and Sun, Liyang},
	title        = {Weak Instruments in Instrumental Variables Regression: Theory and Practice},
	journal      = {Annual Review of Economics},
	volume       = 11,
	number       = 1,
	pages        = {727--753},
	year         = 2019
}

@article{angrist1996identification,
	author       = {Angrist, Joshua D and Imbens, Guido W and Rubin, Donald B},
	journal      = {Journal of the American Statistical Association},
	number       = 434,
	pages        = {444--455},
	publisher    = {Taylor \& Francis},
	title        = {Identification of Causal Effects Using Instrumental Variables},
	volume       = 91,
	year         = 1996
}

@book{angrist2008mostly,
	author       = {Angrist, Joshua D and Pischke, J{\"o}rn-Steffen},
	publisher    = {Princeton University Press},
	title        = {Mostly Harmless Econometrics},
    address={Princeton, NJ},
	year         = 2009
}

@article{arellano2002sargan,
	author       = {Arellano, Manuel},
	journal      = {Journal of Business \& Economic Statistics},
	number       = 4,
	pages        = {450--459},
	publisher    = {Taylor \& Francis},
	title        = {Sargan's Instrumental Variables Estimation and the Generalized Method of Moments},
	volume       = 20,
	year         = 2002
}

@article{arias2019large,
	author       = {Arias, Eric and Stasavage, David},
	journal      = {The Journal of Politics},
	number       = 4,
	pages        = {1517--1522},
	title        = {How Large are the Political Costs of Fiscal Austerity?},
	volume       = 81,
	year         = 2019,
	publisher    = {The University of Chicago Press Chicago, IL}
}

@article{baccini2021,
	title        = {Gone for Good: Deindustrialization, White Voter Backlash, and US Presidential Voting},
	author       = {Baccini, Leonardo and Weymouth, Stephen},
	journal      = {American Political Science Review},
	volume       = 115,
	number       = 2,
	pages        = {550--567},
	year         = 2021,
	publisher    = {Cambridge University Press}
}

@article{bound2000compulsory,
	author       = {Bound, John and Jaeger, David A},
	journal      = {Research in Labor Economics},
	number       = 4,
	pages        = {83--108},
	title        = {Do Compulsory School Attendance Laws Alone Explain the Association between Quarter of Birth and Earnings?},
	volume       = 19,
	year         = 2000
}

@article{bun2010weak,
	author       = {Bun, Maurice JG and Windmeijer, Frank},
	journal      = {The Econometrics Journal},
	number       = 1,
	pages        = {95--126},
	publisher    = {Oxford University Press Oxford, UK},
	title        = {The Weak Instrument Problem of the System GMM Estimator in Dynamic Panel Data Models},
	volume       = 13,
	year         = 2010
}

@article{cameron2008bootstrap,
	author       = {Cameron, A Colin and Gelbach, Jonah B and Miller, Douglas L},
	journal      = {The Review of Economics and Statistics},
	number       = 3,
	pages        = {414--427},
	publisher    = {The MIT Press},
	title        = {Bootstrap-based Improvements for Inference with Clustered Errors},
	volume       = 90,
	year         = 2008
}

@article{chernozhukov2008reduced,
	title        = {The Reduced Form: A Simple Approach to Inference with Weak Instruments},
	author       = {Chernozhukov, Victor and Hansen, Christian},
	journal      = {Economics Letters},
	volume       = 100,
	number       = 1,
	pages        = {68--71},
	year         = 2008,
	publisher    = {Elsevier}
}

@article{cinelli2022omitted,
	title        = {An Omitted Variable Bias Framework for Sensitivity Analysis of Instrumental Variables},
	author       = {Cinelli, Carlos and Hazlett, Chad},
	journal      = {Available at SSRN 4217915},
	year         = 2022
}

@article{Colin_Cameron2015-wp,
	author       = {Colin Cameron, A and Miller, Douglas L},
	journal      = {The Journal of Human Resources},
	number       = 2,
	pages        = {317--372},
	publisher    = {University of Wisconsin Press},
	title        = {A Practitioner's Guide to {Cluster-robust} Inference},
	volume       = 50,
	year         = 2015
}

@article{Conley2012-mu,
	author       = {Conley, Timothy G and Hansen, Christian B and Rossi, Peter E},
	journal      = {The Review of Economics and Statistics},
	number       = 1,
	pages        = {260--272},
	publisher    = {MIT Press},
	title        = {Plausibly Exogenous},
	volume       = 94,
	year         = 2012
}

@article{davidson2015bootstrap,
	author       = {Davidson, Russell and MacKinnon, James G},
	journal      = {Econometrics},
	number       = 4,
	pages        = {825--863},
	publisher    = {Multidisciplinary Digital Publishing Institute},
	title        = {Bootstrap Tests For Overidentification in Linear Regression Models},
	volume       = 3,
	year         = 2015
}

@article{dieterle2016simple,
	author       = {Dieterle, Steven G and Snell, Andy},
	journal      = {Labour Economics},
	pages        = {76--86},
	publisher    = {Elsevier},
	title        = {A Simple Diagnostic to Investigate Instrument Validity and Heterogeneous Effects when using a Single Instrument},
	volume       = 42,
	year         = 2016
}

@article{dinas2014,
	author       = {Dinas, Elias},
	journal      = {American Journal of Political Science},
	number       = 2,
	pages        = {449--465},
	title        = {Does Choice Bring Loyalty? Electoral Participation and the Development of Party Identification},
	volume       = 58,
	year         = 2014
}

@article{dorsch_maarek2019,
	author       = {Dorsch, Michael T. and Maarek, Paul},
	journal      = {American Political Science Review},
	number       = 2,
	pages        = {385--404},
	title        = {Democratization and the Conditional Dynamics of Income Distribution},
	volume       = 113,
	year         = 2019
}

@article{dower_etal2018,
	author       = {Dower, Paul Casta{\~n}eda and Finkel, Evgeny and Gehlbach, Scott and Nafziger, Steven},
	journal      = {American Political Science Review},
	number       = 1,
	pages        = {125--147},
	title        = {Collective Action and Representation in Autocracies: Evidence from Russia's Great Reforms},
	volume       = 112,
	year         = 2018
}

@article{dube2015,
	author       = {{Dube}, {Oeindrila} and {Naidu}, {Suresh}},
	date         = 2015,
	journal      = {The Journal of Politics},
	number       = 1,
	pages        = {249{\textendash}267},
	title        = {Bases, Bullets, and Ballots: The Effect of US Military Aid on Political Conflict in Colombia},
	volume       = 77,
	year         = 2015
}

@article{Esarey2019-qt,
	author       = {Esarey, J and Menger, A},
	journal      = {Political Science Research and Methods},
	number       = 3,
	pages        = {541--559},
	title        = {Practical and Effective Approaches to Dealing with Clustered Data},
	volume       = 7,
	year         = 2019
}

@article{FeltonStewart2022,
	author       = {Felton, Chris and Stewart, Brandon M.},
	title        = {Handle with Care: A Sociologist's Guide to Causal Inference with Instrumental Variables},
	year         = 2022,
url={https://osf.io/preprints/socarxiv/3ua7q}

}

@article{fieller1954some,
	author       = {Fieller, Edgar C},
	journal      = {Journal of the Royal Statistical Society: Series B (Methodological)},
	number       = 2,
	pages        = {175--185},
	title        = {Some Problems in Interval Estimation},
	volume       = 16,
	year         = 1954
}

@article{gelman2014beyond,
	author       = {Gelman, Andrew and Carlin, John},
	journal      = {Perspectives on Psychological Science},
	number       = 6,
	pages        = {641--651},
	publisher    = {Sage Publications Sage CA: Los Angeles, CA},
	title        = {Beyond Power Calculations: Assessing Type S (sign) and Type M (magnitude) Errors},
	volume       = 9,
	year         = 2014
}

@book{Gerber2012-fr,
	address      = {New York, NY},
	author       = {Gerber, Alan S and Green, Donald P},
	publisher    = {W. W. Northon},
	title        = {Field Experiments: Design, Analysis and Interpretation},
	year         = 2012
}

@book{greene2003econometric,
	author       = {Greene, William H},
	publisher    = {Pearson Education India},
	title        = {Econometric Analysis},
    address={Noida, India},
	year         = 2003
}

@article{hager2019,
	author       = {{Hager}, {Anselm} and {Hilbig}, {Hanno}},
	date         = 2019,
	journal      = {American Journal of Political Science},
	number       = 4,
	pages        = {758{\textendash}773},
	title        = {Do Inheritance Customs Affect Political and Social Inequality?},
	volume       = 63,
	year         = 2019
}

@article{Hahn2021-lq,
	author       = {Hahn, Jinyong and Liao, Zhipeng},
	journal      = {Econometrica},
	number       = 4,
	pages        = {1963--1977},
	publisher    = {The Econometric Society},
	title        = {Bootstrap Standard Error Estimates and Inference},
	volume       = 89,
	year         = 2021
}

@article{Hainmueller2019-wx,
	title        = {How Much Should We Trust Estimates from Multiplicative Interaction Models? Simple Tools to Improve Empirical Practice},
	volume       = 27,
	number       = 2,
	journal      = {Political Analysis},
	author       = {Hainmueller, Jens and Mummolo, Jonathan and Xu, Yiqing},
	year         = 2019,
	pages        = {163–192}
}

@article{hall1996bootstrap,
	title        = {Bootstrap Critical Values for Tests Based on Generalized-method-of-moments Estimators},
	author       = {Hall, Peter and Horowitz, Joel L},
	journal      = {Econometrica},
	pages        = {891--916},
	year         = 1996,
	volume       = 64,
	number       = 4,
	publisher    = {Jstor}
}

@article{hansen1982large,
	author       = {Hansen, Lars Peter},
	journal      = {Econometrica},
	pages        = {1029--1054},
	publisher    = {Jstor},
	title        = {Large Sample Properties of Generalized Method of Moments Estimators},
	volume       = 50,
	number       = 4,
	year         = 1982
}

@article{henderson2016mediating,
	author       = {Henderson, John and Brooks, John},
	journal      = {The Journal of Politics},
	number       = 3,
	pages        = {653--669},
	title        = {Mediating the Electoral Connection: The Information Effects of Voter Signals on Legislative Behavior},
	volume       = 78,
	year         = 2016,
	publisher    = {University of Chicago Press Chicago, IL}
}

@article{hirano2015location,
	title        = {Location Properties of Point Estimators in Linear Instrumental Variables and Related Models},
	author       = {Hirano, Keisuke and Porter, Jack R},
	journal      = {Econometric Reviews},
	volume       = 34,
	number       = {6-10},
	pages        = {720--733},
	year         = 2015,
	publisher    = {Taylor \& Francis}
}

@article{Ishimaru2021-ik,
	archiveprefix = {arXiv},
	author       = {Ishimaru, Shoya},
	title        = {Empirical Decomposition of The {IV-OLS} Gap with Heterogeneous and Nonlinear Effects},
	url          = {http://arxiv.org/abs/2101.04346},
	year         = 2021
}

@article{jiang2017have,
	author       = {Jiang, Wei},
	journal      = {The Review of Corporate Finance Studies},
	number       = 2,
	pages        = {127--140},
	publisher    = {Oxford University Press},
	title        = {Have Instrumental Variables Brought Us Closer to the Truth},
	volume       = 6,
	year         = 2017
}

@article{key2016we,
	author       = {Key, Ellen M},
	journal      = {PS: Political Science \& Politics},
	number       = 2,
	pages        = {268--272},
	publisher    = {Cambridge University Press},
	title        = {How are We Doing? Data Access and Replication in Political Science},
	volume       = 49,
	year         = 2016
}

@article{kim2019,
	author       = {{Kim}, {Jeong Hyun}},
	date         = 2019,
	journal      = {American Journal of Political Science},
	number       = 3,
	pages        = {594{\textendash}610},
	title        = {Direct Democracy and Women's Political Engagement},
	volume       = 63,
	year         = 2019
}

@article{Lee2020-mi,
	title        = {Valid t-ratio Inference for IV},
	author       = {Lee, David S and McCrary, Justin and Moreira, Marcelo J and Porter, Jack},
	journal      = {American Economic Review},
	volume       = 112,
	number       = 10,
	pages        = {3260--90},
	year         = 2022
}

@article{lorentzen_etal2014,
	author       = {Lorentzen, Peter and Landry, Pierre and Yasuda, John},
	journal      = {The Journal of Politics},
	number       = 1,
	pages        = {182--194},
	title        = {Undermining Authoritarian Innovation: The Power of China's Industrial Giants},
	volume       = 76,
	year         = 2014
}

@article{Marbach2020-ts,
	author       = {Marbach, Moritz and Hangartner, Dominik},
	journal      = {Political Analysis},
	number       = 3,
	pages        = {435--444},
	publisher    = {Cambridge University Press},
	title        = {Profiling Compliers and Noncompliers for {Instrumental-Variable} Analysis},
	volume       = 28,
	year         = 2020
}

@article{mellon2020rain,
	author       = {Mellon, Jonathan},
	url      = {10.31235/osf.io/9qj4f},
	title        = {Rain, Rain, Go Away: 195 Potential Exclusion-restriction Violations for Studies using Weather as an Instrumental Variable},
	year         = 2023
}

@article{Mikusheva2006-lk,
	author       = {Mikusheva, Anna and Poi, Brian P},
	journal      = {The Stata Journal},
	number       = 3,
	pages        = {335--347},
	publisher    = {SAGE Publications},
	title        = {Tests and Confidence Sets with Correct Size when Instruments are Potentially Weak},
	volume       = 6,
	year         = 2006
}

@article{Moreira2003-oj,
	author       = {Moreira, Marcelo J},
	journal      = {Econometrica},
	number       = 4,
	pages        = {1027--1048},
	publisher    = {The Econometric Society},
	title        = {A Conditional Likelihood Ratio Test For Structural Models},
	volume       = 71,
	year         = 2003
}

@article{moreira2009tests,
	title        = {Tests with Correct Size when Instruments can be Arbitrarily Weak},
	author       = {Moreira, Marcelo J},
	journal      = {Journal of Econometrics},
	volume       = 152,
	number       = 2,
	pages        = {131--140},
	year         = 2009,
	publisher    = {Elsevier}
}

@article{nelson1990some,
	author       = {Nelson, Charles and Starz, Richard},
	journal      = {Econometrica},
	number       = 41,
	pages        = {967--976},
	title        = {Some Further Results on the Exact Small Sample Properties of the Instrumental Variables Estimator},
	volume       = 58,
	year         = 1990
}

@article{Olea2013-pa,
	author       = {Olea, Jos{\'e} Luis Montiel and Pflueger, Carolin},
	journal      = {Journal of Business \& Economic Statistics},
	number       = 3,
	pages        = {358--369},
	title        = {A Robust Test for Weak Instruments},
	volume       = 31,
	year         = 2013
}

@article{rueda2017,
	title        = {Small Aggregates, Big Manipulation: Vote Buying Enforcement and Collective Monitoring},
	author       = {{Rueda}, {Miguel R.}},
	year         = 2017,
	date         = 2017,
	journal      = {American Journal of Political Science},
	pages        = {163--177},
	volume       = 61,
	number       = 1
}

@article{sekhon2012natural,
	author       = {Sekhon, Jasjeet S and Titiunik, Rocio},
	journal      = {American Political Science Review},
	pages        = {35--57},
	publisher    = {Jstor},
	number       = 1,
	title        = {When Natural Experiments are neither Natural nor Experiments},
	year         = 2012
}

@article{sovey2011instrumental,
	author       = {Sovey, Allison J and Green, Donald P},
	journal      = {American Journal of Political Science},
	number       = 1,
	pages        = {188--200},
	title        = {Instrumental Variables Estimation in Political Science: A Readers' Guide},
	volume       = 55,
	year         = 2011
}

@article{Staiger1997-lo,
	author       = {Staiger, Douglas and Stock, James H},
	journal      = {Econometrica},
	number       = 3,
	pages        = {557--586},
	title        = {Instrumental Variables Regression with Weak Instruments},
	volume       = 65,
	year         = 1997
}

@article{stock2002survey,
	title        = {A Survey of Weak Instruments and Weak Identification in Generalized Method of Moments},
	author       = {Stock, James H and Wright, Jonathan H and Yogo, Motohiro},
	journal      = {Journal of Business \& Economic Statistics},
	volume       = 20,
	number       = 4,
	pages        = {518--529},
	year         = 2002
}

@inbook{stock2005asymptotic,
	adderess        = {Cambridge},
	title        = {Asymptotic Distributions of Instrumental Variables Statistics with Many Instruments},
	booktitle    = {Identification and Inference for Econometric Models: Essays in Honor of Thomas Rothenberg},
	publisher    = {Cambridge University Press},
	author       = {Stock, James H. and Yogo, Motohiro},
	editor       = {Andrews, Donald W. K. and Stock, James H.Editors},
	year         = 2005,
	pages        = {109–120}
}

@article{vernby2013,
	author       = {Vernby, Kare},
	journal      = {American Journal of Political Science},
	number       = 1,
	pages        = {15--29},
	title        = {Inclusion and Public Policy: Evidence from Sweden's Introduction of Noncitizen Suffrage},
	volume       = 57,
	year         = 2013,
	date         = 2013
}

@article{Young2022,
	author       = {Young, Alwyn},
	journal      = {European Economic Review},
	title        = {Consistency without Inference: Instrumental Variables in Practical Application},
	volume       = 147,
	pages        = {104--112},
	year         = 2022
}

@article{zhu2017,
	author       = {Zhu, Boliang},
	journal      = {American Journal of Political Science},
	number       = 1,
	pages        = {84--99},
	title        = {{MNCs}, Rents, and Corruption: Evidence from China},
	volume       = 61,
	year         = 2017,
	date         = 2017
}

@article{alt2016,
  title={Credible Sources and Sophisticated Voters: When does New Information Induce Economic Voting?},
  author={Alt, James E and Marshall, John and Lassen, David D},
  journal={The Journal of Politics},
  volume={78},
  number={2},
  pages={327--342},
  year={2016},
  publisher={University of Chicago Press Chicago, IL}
}

@article{betz2018use,
  title={On the Use and Abuse of Spatial Instruments},
  author={Betz, Timm and Cook, Scott J and Hollenbach, Florian M},
  journal={Political Analysis},
  volume={26},
  number={4},
  pages={474--479},
  year={2018},
  publisher={Cambridge University Press}
}

\end{document}